\documentclass[twoside,english,3p]{elsarticle}
\usepackage[T1]{fontenc}
\usepackage[latin9]{inputenc}
\usepackage{amsmath}
\usepackage{graphicx}
\usepackage{algorithm}
\usepackage[noend]{algpseudocode}

\makeatletter

\newcommand{\noun}[1]{\textsc{#1}}

\journal{}

\makeatother

\usepackage{babel}
\begin{document}

\begin{frontmatter}{}

\title{Exact and Locally Implicit Source Term Solvers for Multifluid-Maxwell Systems}

\author[pu,pppl]{Liang Wang\corref{cor1}}
\ead{lwang@pppl.gov}

\author[pppl]{Ammar H. Hakim}

\author[umcp,pu,pppl,gsfc]{Jonathan Ng}

\author[pu,pppl]{Chuanfei Dong}

\author[unh]{Kai Germaschewski}

\address[pu]{Princeton University, Princeton, NJ, 08544}

\address[pppl]{Princeton Plasma Physics Laboratory, Princeton, NJ 08543}

\address[umcp]{University of Maryland, College Park, College Park, MD 20742}

\address[gsfc]{NASA Goddard Space Flight Center, Greenbelt, MD 20771}

\address[unh]{Space Science Center, University of New Hampshire, Durham, NH 03824}

\cortext[cor1]{Corresponding author.}
\begin{abstract}
Recently, a family of models that couple multifluid systems to the full Maxwell equations have been used in laboratory, space, and astrophysical plasma modeling. These models are more complete descriptions of the plasma than reduced models like magnetohydrodynamic (MHD) since they 
are derived more closely from the full kinetic Vlasov-Maxwell system, without assumptions like quasi-neutrality, negligible electron mass, etc. Thus these models
naturally retain non-ideal MHD effects like electron inertia, Hall term, pressure anisotropy/nongyrotropy, displacement current, among others. One obstacle to broader application of these model is that an explicit treatment of their source terms leads to the need to resolve rapid processes like plasma oscillation and electron cyclotron motion, even when these are not important. In this paper, we suggest two ways to address this issue. First, we derive the analytic solutions to the source update equations, which can be implemented as a practical, but less generic solver. We then develop a time-centered, locally implicit algorithm to update the source terms, allowing stepping over the fast kinetic time-scales. For a plasma with $S$ species, the locally implicit algorithm involves inverting a local $(3S+3)\times(3S+3)$ matrix only, thus is very efficient. The performance can be further increased by using the direct update formulas to skip null calculations. We present benchmarks illustrating the exact energy-conservation of the locally implicit solver, as well as its efficiency and robustness for both small-scale, idealized problems and large-scale, complex systems. The locally implicit algorithm can be also easily extended to include other local sources, like collisions and ionization, which are difficult to solve analytically.
\end{abstract}

\end{frontmatter}{}


\section{Introduction}
In this paper we describe numerical methods to update source terms
for the multifluid plasma equations coupled to Maxwell equations.
These models consist of equations of 

velocity moments for each plasma species
$s$, in the conservative form,
\begin{eqnarray}
\frac{\partial_{t}\left(m_{s}n_{s}\right)}{\partial t}+\frac{\partial\left(m_{s}n_{s}u_{j,s}\right)}{\partial x_{j}} & = & 0\label{eq:drhodt_full}\\
\frac{\partial\left(m_{s}n_{s}u_{j,s}\right)}{\partial t}+\frac{\partial\mathcal{P}_{ij,s}}{\partial x_{j}} & = & n_{s}q_{s}\left(E_{i}+\epsilon_{ijk}u_{j,s}B_{k}\right)\label{eq:drhovdt_full}
\end{eqnarray}
Here, $q_{s}$ and $m_{s}$ are the particle charge and mass, $\varepsilon_{ijk}$
is the Levi-Civita symbol. The moments are defined as 
\begin{align}
n_{s}\left(\mathbf{x}\right) & \equiv\int f_{s}d\mathbf{v}\\
m_{s}n_{s}\left(\mathbf{x}\right)u_{i,s}\left(\mathbf{x}\right) & \equiv m_{s}\int v_{i}f_{s}d\mathbf{v}\\
\mathcal{P}_{ij,s}\left(\mathbf{x}\right) & \equiv m_{s}\int v_{i}v_{j}f_{s}d\mathbf{v}
\end{align}
with $f_{s}\left(\mathbf{x},\mathbf{v},t\right)$ being the phase
space distribution function. We will neglect the subscript $s$ hereinafter
for convenience. For completeness, $\mathcal{P}_{ij}$ relates to
the more familiar thermal pressure tensor 
\begin{equation}
P_{ij}\equiv m\int\left(v_{i}-u_{i}\right)\left(v_{j}-u_{j}\right)fd\mathbf{v}
\end{equation}
 by 
\begin{equation}
\mathcal{P}_{ij}=P_{ij}+nmu_{i}u_{j}.
\end{equation}
For simplicity, non-ideal effects like viscous
dissipation are neglected. The electric and magnetic fields $\mathbf{E}$
and $\mathbf{B}$ are evolved using Maxwell equations 
\begin{align}
\frac{\partial\mathbf{B}}{\partial t}+\nabla\times\mathbf{E} & =0\label{eq:dBdt_full}\\
\frac{\partial\mathbf{E}}{\partial t}-c^{2}\nabla\times\mathbf{B} & =-\frac{1}{\varepsilon_{0}}\sum_{s}q_{s}n_{s}\mathbf{u}_{s}.\label{eq:dEdt_full}
\end{align}
with $c=1/\sqrt{\mu_{0}\varepsilon_{0}}$ being the speed of light.

To close the system, the second order moment $\mathcal{P}_{ij}$ or $P_{ij}$
must be specified. For example, a cold fluid closure simply sets $P_{ij}=0$,
while an isothermal equation of state (EOS) assumes that the temperature
is constant. Or, assuming zero heat flux and that the pressure tensor
is isotropic, we can write an adiabatic EOS for $P_{ij}=p\boldsymbol{\mathcal{I}}$
\begin{align}
\frac{\partial\mathcal{E}}{\partial t}+\nabla\cdot\left[\left(p+\mathcal{E}\right)\mathbf{u}\right] & =nq\mathbf{u}\cdot\mathbf{E},\label{eq:dpdt_full}
\end{align}
where 
\begin{equation}
\mathcal{E}\equiv\frac{p}{\gamma-1}+\frac{1}{2}\rho\left|\mathbf{u}\right|^{2}
\end{equation}
is the total fluid (thermal plus kinetic) energy and $\gamma$ is
the adiabatic index, set to $5/3$ for a fully ionized plasma. For
a plasma with $S$ species ($s=1,\ldots,S$) this system is closed
and has a total of $5S+6$ equations, and are here referred to as
the \noun{five-moment} model \citep{Hakim2006}. More general models
can be obtained by retaining the evolution equations for all six components
of the pressure tensor\citep{Hakim2008,Johnson2014} in the so-called
\noun{ten-moment} model
\begin{equation}
\frac{\partial\mathcal{P}_{ij,s}}{\partial t}+\frac{\partial\mathcal{Q}_{ijm,s}}{\partial x_{m}}=n_{s}q_{s}u_{[i,s}E_{j]}+\frac{q_{s}}{m_{s}}\epsilon_{[iml}\mathcal{P}_{mj],s}B_{l}\label{eq:dPdt_full}
\end{equation}
where the third moment 
\begin{equation}
\mathcal{Q}_{ijm,s}\left(\mathbf{x}\right)\equiv m_{s}\int v_{i}v_{j}v_{m}f_{s}d\mathbf{v}
\end{equation}
relates to the heat flux tensor defined in the fluid frame
\begin{equation}
Q_{ijm}\equiv m\int(v_{i}-u_{i})(v_{j}-u_{j})(v_{m}-u_{m})fd\mathbf{v}
\end{equation}
 by 
\begin{equation}
\mathcal{Q}_{ijm}=Q_{ijm}+u_{[i}\mathcal{P}_{jm]}-2nmu_{i}u_{j}u_{m}.
\end{equation}
Again, the equations here must be closed by some approximation for
the heat-flux tensor. Another option is to include evolution equations for even
higher order moments, e.g., the ten independent components of the heat-flux
tensor\citep{Ng2015a}.

Although multifluid-Maxwell models provide a more complete description
of the plasma than reduced, asymptotic models like magnetohydrodynamics
(MHD) \citep{Srinivasan2011,Wang2015}, they are less frequently used.
The reason for this is the fast kinetic scales involved.
Retaining the electron inertia adds plasma-frequency and cyclotron
time-scale, while non-neutrality adds Debye length spatial-scales.
Further, inclusion of the displacement currents means that electromagnetic
(EM) waves must be resolved when using an explicit scheme. Fortunately,
the restrictions due to kinetic scales are introduced only through
the non-hyperbolic source terms of Eqns.~(\ref{eq:drhovdt_full}),
(\ref{eq:dEdt_full}), and Eqn.~(\ref{eq:dPdt_full}). Therefore
we may eliminate these restrictions by updating the source term separately
either exactly or using an implicit algorithm. This allows larger 
time steps and leads to significant
speedup, especially with realistic electron/ion mass ratios. Developing such
source term update schemes is the focus of this paper. The speed of
light constraint still exists, however, can be greatly relaxed, using
reduced values for the speed of light and/or sub-cycling
Maxwell equations. Of course, an implicit Maxwell solver, or a reduced
set of electromagnetic equations like the Darwin approximation\citep{birdsallbook},
can also relax the time-step restrictions. In either case, though,
a fully implicit approach is needed, which is not considered in this
paper.

The rest of the paper is organized as follows. First, the source term
update equations are written down as time-dependent constant-coefficient
ordinary differential equations (ODEs). We then give exact solutions
to these equations for any number of plasma species. Subsequently,
a locally implicit algorithm is presented. It is shown that the time-steps
are restricted solely by the speed of light, and that the algorithm
preserves positivity of density and pressure. The accuracy and robustness
of both (i.e., the analytic solution and the locally implicit solution)
methods are demonstrated through a few standard benchmark problems,
as well as through an application to large-scale modeling of the interaction
between solar wind and Earth's magnetosphere. The appendix gives a
thorough derivation of the exact solutions plus direct formulae for
the locally implicit schemes. The eigensystem of the ten-moment model,
useful for implementing approximate Riemann solvers for this system,
is provided in the appendix too.

\section{An Operator Splitting Scheme and The Source Term Update Equations}
The multifluid-Maxwell equations can be split into a homogeneous
part and a source term update part. The key idea is to solve these
two parts separately and apply high accuracy schemes on both.

Ignoring sources, the \emph{homogeneous} equations can be solved in
the \emph{conservation law form
\begin{equation}
\frac{\partial\mathbf{Q}}{\partial t}+\nabla\cdot\mathbf{F}=0\label{eq:homosys}
\end{equation}
}where $\mathbf{Q}$ is the vector of conserved quantities (fluid
moments and electromagnetic field) and $\mathbf{F}$ are the corresponding
fluxes. See Eqns.\,(1)\textendash (3) of \citep{Hakim2006} for the
conservation form of five-moment equations, and Eqns.\,(24)\textendash (25)
of \citep{Hakim2008} for the conservation form of the ten-moment
equations. An explicit solution of Eqn.~(\ref{eq:homosys}) is subject
to light speed constraint $c\Delta t/\Delta x<\mathrm{CFL}$, where
$\Delta x$ is the (smallest) grid spacing and $\mathrm{CFL}\leq1$
is determined from the spatial scheme used. The often more restrictive
constraints due to kinetic scales are contained in the remaining non-hyperbolic,
source term update part only. This implies that it is possible to
eliminate these constraints if a proper implicit algorithm is applied
to update the source term.

Deferring the source term details to the next paragraph, we represent the
\emph{homogeneous} update schematically as the operator
$\exp\left(\mathcal{L}_{H}\Delta t\right)$ and the \emph{source}
update as $\exp\left(\mathcal{L}_{S}\Delta t\right)$. The full algorithm
can now be written as the Strang-splitting sequence that has second
order accuracy in time\citep{Strang1968},
\begin{equation}
\exp\left(\mathcal{L}_{S}\Delta t/2\right)\exp\left(\mathcal{L}_{H}\Delta t\right)\exp\left(\mathcal{L}_{S}\Delta t/2\right).
\end{equation}
The remainder of this paper is devoted to developing schemes for $\exp\left(\mathcal{L}_{S}\Delta t\right)$.
As for $\exp\left(\mathcal{L}_{H}\Delta t\right)$, one can use a
number of schemes, including the finite-volume (FV) wave-propagation scheme\citep{LeVeque2002,Hakim2006},
a variation of the MUSCL algorithm\citep{kulikovskii_2001}, or
a discontinuous Galerkin (DG) scheme\citep{Cockburn:2001vr,Hesthaven2008,Loverich:2010ea},
among others. Each of these schemes has advantages: finite-volume
methods are robust and easy to implement, while DG schemes are high-order
and have the potential of providing higher accuracy at a lower cost
compared to second-order schemes.


Now we consider the source update equations. For the five-moment model, the source terms are 
\begin{equation}
\begin{cases}
\frac{\partial\mathbf{J}_{s}}{\partial t} & =\omega_{s}^{2}\varepsilon_{0}\mathbf{E}+\mathbf{J}_{s}\times\boldsymbol{\Omega}_{s}\\
\varepsilon_{0}\frac{\partial\mathbf{E}}{\partial t} & =-\sum_{s}\mathbf{J}_{s}.
\end{cases}\label{eq:dJdt_dEdt_source}
\end{equation}
Here, $\boldsymbol{\Omega}_{s}\equiv q_{s}\mathbf{B}/m_{s}$ is the
cyclotron frequency, $\omega_{s}\equiv\sqrt{q_{s}^{2}n_{s}/\varepsilon_{0}m_{s}}$
is the species plasma frequency, and we use currents, $\mathbf{J}_{s}\equiv q_{s}n_{s}\mathbf{u}_{s}$,
instead of momentum for convenience. During the source term update,
the plasma density $n_{s}$ and magnetic field $\mathbf{B}$ remain
unchanged and this coupled system has an energy invariant, 
\begin{equation}
\sum_{s}\frac{1}{2}\frac{\mathbf{J}_{s}^{2}}{\varepsilon_{0}\omega_{s}^{2}}+\frac{\varepsilon_{0}}{2}\mathbf{E}^{2}=\mathrm{const}.\label{eq:src-energy}
\end{equation}

The ten-moment model has the same source terms for currents and the
electric field. In addition, there are source terms in the pressure
tensor equation (\ref{eq:dPdt_full}) that accounts for a rotation
around the background magnetic field
\begin{equation}
\frac{\partial\mathbf{P}_{s}}{\partial t}={\rm Sym2}\left(\mathbf{P}_{s}\times\boldsymbol{\Omega}_{s}\right).\label{eq:dPdt_source}
\end{equation}
Here, ${\rm Sym2\left(\mathbf{V}\right)}$ denotes the space of all
symmetric tensors of 2nd order defined on the tensor $\mathbf{V}$.

\section{\label{sec:exact}An Exact Solution Scheme}

During the source term updates Eqn.~(\ref{eq:dJdt_dEdt_source})
and (\ref{eq:dPdt-original}), the plasma densities and magnetic field
remain unchanged. Consequently, the equations are effectively constant-coefficient
linear ODEs in time. They can be solved exactly to obtain currents/electric
field or pressure tensor as functions of time. Ref.~\citep{Johnson2014}
obtained such exact, analytic solutions in the two-fluid case. Here,
we unify and generalize the procedure for any number of species and
outline the critical steps/results.

\subsection{Exact Solutions of the Electric Field-Currents Source Term Updates}

For a single grid cell, the procedure to update the source term Eqn.~(\ref{eq:dJdt_dEdt_source}) by a time step $\Delta t$ is outlined below:

\begin{algorithm}
\caption{Update source term Eqn.~(\ref{eq:dJdt_dEdt_source}) exactly in one cell}\label{alg:exact-one-cell}
\begin{algorithmic}[1]
\State Normalize $\mathbf{E}$ and $\mathbf{J}_{s}$ to get anti-symmetric;\label{lst:line:exact-norm}
coefficient matrix for Eqn.~(\ref{eq:dJdt_dEdt_source})
\If {If $\left|\mathbf{B}\right|\neq0$}
  \State Decompose the system into a parallel part and a perpendicular part regarding $\mathbf{B}$; \label{lst:line:exact-decomp}
  \For{sub-system in \{parallel, perpendicular\}} \label{lst:line:exact-for-par-perp}
    \State Compute eigenvalues and eigenvectors of the coefficient matrix at $t=0$;
    \State Compute eigencoefficients by projecting the initial state onto the eigenvectors;
    \State Compute eigenvectors at $t=\Delta t$;
    \State Compute the updated state at $t=\Delta t$ as combinations of the updated eigenvectors;
  \EndFor
\Else
  \State Solve the equivalent parallel problem along all three directions;
\EndIf
\State Normalize $\mathbf{E}$ and $\mathbf{J}_{s}$ back to their original units;
\end{algorithmic}
\end{algorithm}

Here we briefly reiterate some of the steps and give the results but
leave the more involved details to App.~\ref{sec:appendix-exact}.

\subsubsection{Normalization Towards an Anti-Symmetric System}

This section addresses the line~\ref{lst:line:exact-norm} in Algorithm~(\ref{alg:exact-one-cell}).
Eqn.~(\ref{eq:dJdt_dEdt_source}) for $N$ species can be written
in a matrix form 
\begin{equation}
\frac{\partial}{\partial t}\left[\begin{array}{c}
\mathbf{E}\\
\mathbf{J}_{1}\\
\mathbf{J}_{2}\\
\vdots\\
\mathbf{J}_{N}
\end{array}\right]=\left[\begin{array}{ccccc}
0 & -1/\varepsilon_{0} & -1/\varepsilon_{0} & \cdots & -1/\varepsilon_{0}\\
\omega_{1}^{2}\varepsilon_{0} & -\boldsymbol{\Omega}_{1}\times\boldsymbol{\mathcal{I}} & 0 & \cdots & 0\\
\omega_{2}^{2}\varepsilon_{0} & 0 & -\boldsymbol{\Omega}_{2}\times\boldsymbol{\mathcal{I}} & \cdots & 0\\
\vdots & \vdots & \vdots & \ddots & \vdots\\
\omega_{N}^{2}\varepsilon_{0} & 0 & 0 & \cdots & -\boldsymbol{\Omega}_{N}\times\boldsymbol{\mathcal{I}}
\end{array}\right]\left[\begin{array}{c}
\mathbf{E}\\
\mathbf{J}_{1}\\
\mathbf{J}_{2}\\
\vdots\\
\mathbf{J}_{N-1}\\
\mathbf{J}_{N}
\end{array}\right],\label{eq:dE-dJ-matrix}
\end{equation}
where $\boldsymbol{\mathcal{I}}$ is a $3\times3$ unit tensor, $\omega_{s}$
and $\boldsymbol{\Omega}_{s}$ denote the plasma and signed cyclotron
frequency of species $s$
\begin{equation}
\omega_{s}\equiv\sqrt{\frac{q_{s}^{2}n_{s}}{\varepsilon_{0}m_{s}}},\quad\boldsymbol{\Omega}_{s}\equiv\frac{q_{s}\mathbf{B}}{m_{s}}.
\end{equation}
It is easier to work with a symmetric or antisymmetric system. This
can be achieved by renormaliznig the electric field and currents.
Normalizating $E_{0}$ and $J_{s0}$ so that
\begin{equation}
\mathbf{E}=\tilde{\mathbf{E}}E_{0},\quad\mathbf{J}_{s}=\tilde{\mathbf{J}}_{s}J_{s0},\label{eq:Enorm-Jnorm}
\end{equation}
and require 
\begin{equation}
J_{s0}/\varepsilon_{0}E_{0}=\omega_{ps},\label{eq:Js0-E0-relation}
\end{equation}
the system becomes anti-symmetric:
\begin{equation}
\frac{\partial}{\partial t}\left[\begin{array}{c}
\tilde{\mathbf{E}}\\
\tilde{\mathbf{J}}_{1}\\
\tilde{\mathbf{J}}_{2}\\
\vdots\\
\tilde{\mathbf{J}}_{N}
\end{array}\right]=\left[\begin{array}{ccccc}
0 & -\omega_{1} & -\omega_{2} & \cdots & -\omega_{N}\\
\omega_{1} & -\boldsymbol{\Omega}_{1}\times\boldsymbol{\mathcal{I}} & 0 & \cdots & 0\\
\omega_{2} & 0 & -\boldsymbol{\Omega}_{2}\times\boldsymbol{\mathcal{I}} & \cdots & 0\\
\vdots & \vdots & \vdots & \ddots & \vdots\\
\omega_{N} & 0 & 0 & \cdots & -\boldsymbol{\Omega}_{N}\times\boldsymbol{\mathcal{I}}
\end{array}\right]\left[\begin{array}{c}
\tilde{\mathbf{E}}\\
\tilde{\mathbf{J}}_{1}\\
\tilde{\mathbf{J}}_{2}\\
\vdots\\
\tilde{\mathbf{J}}_{N}
\end{array}\right].\label{eq:dE-dJ-matrix-symmetrized}
\end{equation}
Eqn.~(\ref{eq:dE-dJ-matrix-symmetrized}) can be further decomposed
into a parallel part and a perpendicular part with regard to the background
magnetic field direction. We shall solve these two parts separately. 

\subsubsection{Parallel Sub-System}

This section identifies the eigenvectors needed in the loop~\ref{lst:line:exact-for-par-perp} of Algorithm~(\ref{alg:exact-one-cell})
for the \emph{parallel} sub-system. This sub-system evolves $\left(\mathbf{E}_{\parallel};\mathbf{u}_{s\parallel}\right)$
and writes
\begin{equation}
\frac{\partial}{\partial t}\left[\begin{array}{c}
\tilde{\mathbf{E}}_{\parallel}\\
\tilde{\mathbf{J}}_{1\parallel}\\
\tilde{\mathbf{J}}_{2\parallel}\\
\vdots\\
\tilde{\mathbf{J}}_{N\parallel}
\end{array}\right]=\mathbf{M}_{\parallel}\left[\begin{array}{c}
\tilde{\mathbf{E}}_{\parallel}\\
\tilde{\mathbf{J}}_{1\parallel}\\
\tilde{\mathbf{J}}_{2\parallel}\\
\vdots\\
\tilde{\mathbf{J}}_{N\parallel}
\end{array}\right],\label{eq:dE-dJ-para-matrix-form}
\end{equation}
with the coefficient matrix
\begin{equation}
\mathbf{M}_{\parallel}=\left[\begin{array}{ccccc}
0 & -\omega_{1} & -\omega_{2} & \cdots & -\omega_{N}\\
\omega_{1} & 0 & 0 & \cdots & 0\\
\omega_{2} & 0 & 0 & \cdots & 0\\
\vdots & \vdots & \vdots & \ddots & \vdots\\
\omega_{N} & 0 & 0 & \cdots & 0
\end{array}\right].\label{eq:para-mat}
\end{equation}

$\mathbf{M}_{\parallel}$ has three distinct eigenvalues
$
-i\,\omega_{p},\,i\,\omega_{p},\,0
$
with multiplicities 1, 1, and $N-1$, respectively. Here, the total
plasma frequency $\omega_{p}$ is defined as 
\[
\omega_{p}^{2}\equiv\sum_{s}\omega_{s}^{2}.
\]
Following the procedure in Sec.~\ref{subsec:appendix-skew-hermitian-eigenvalue-problem}
and \ref{subsec:appendix-skew-hermitian-real-solutions}, the $N+1$
real solution basis is
\begin{equation}
\left[\begin{array}{c}
\tilde{\mathbf{E}}_{\parallel}\\
\tilde{\mathbf{J}}_{1\parallel}\\
\tilde{\mathbf{J}}_{2\parallel}\\
\vdots\\
\tilde{\mathbf{J}}_{N-1,\parallel}
\tilde{\mathbf{J}}_{N\parallel}
\end{array}\right]=\left[\begin{array}{c}
\omega_{p}\cos\left(\omega_{p}t\right)\\
\omega_{1}\sin\left(\omega_{p}t\right)\\
\omega_{2}\sin\left(\omega_{p}t\right)\\
\vdots\\
\omega_{N-1}\sin\left(\omega_{p}t\right)\\
\omega_{N}\sin\left(\omega_{p}t\right)
\end{array}\right];\left[\begin{array}{c}
-\omega_{p}\sin\left(\omega_{p}t\right)\\
\omega_{1}\cos\left(\omega_{p}t\right)\\
\omega_{2}\cos\left(\omega_{p}t\right)\\
\vdots\\
\omega_{N-1}\cos\left(\omega_{p}t\right)\\
\omega_{N}\cos\left(\omega_{p}t\right)
\end{array}\right];\left[\begin{array}{c}
0\\
1/\omega_{1}\\
0\\
\vdots\\
0\\
-1/\omega_{N}
\end{array}\right],\left[\begin{array}{c}
0\\
0\\
1/\omega_{2}\\
\vdots\\
0\\
-1/\omega_{N}
\end{array}\right],\dots,\left[\begin{array}{c}
0\\
0\\
0\\
\vdots\\
1/\omega_{N-1}\\
-1/\omega_{N}
\end{array}\right].\label{eq:exact-para-eigenvectors-at-t}
    \end{equation}
At $t=0$, they are
\begin{equation}
\left[\begin{array}{c}
\omega_{p}\\
0\\
0\\
\vdots\\
0\\
0
\end{array}\right],\left[\begin{array}{c}
0\\
\omega_{1}\\
\omega_{2}\\
\vdots\\
\omega_{N-1}\\
\omega_{N}
\end{array}\right],\left[\begin{array}{c}
0\\
1/\omega_{1}\\
0\\
\vdots\\
0\\
-1/\omega_{N}
\end{array}\right],\left[\begin{array}{c}
0\\
0\\
1/\omega_{2}\\
\vdots\\
0\\
-1/\omega_{N}
\end{array}\right],\dots,\left[\begin{array}{c}
0\\
0\\
0\\
\vdots\\
1/\omega_{N-1}\\
-1/\omega_{N}
\end{array}\right].\label{eq:exact-para-eigenvectors-at-t-equal-0}
\end{equation}

Note that the $N-1$ degenerate eigenvectors for $\omega_{\parallel}=0$
are not orthogonal when $N\ge3$. Thus we cannot compute their eigencoefficients
following Eqn.~(\ref{eq:ode2matrix-project-to-eigenvectors-how-to})
directly. Instead, we would have to orthogonalize these eigenvectors
(using, e.g., the Gram-Schmidt procedure) first. However, we note
that these eigenvectors do not change in time. As a result, we do
not need to project onto them. The solution at $t$ can
be computed by simply accumulating the incremental changes due to the first
two basis vectors only. In other words, we only need to do projection onto
the first and the second basis vectors in Eqn.~(\ref{eq:exact-para-eigenvectors-at-t-equal-0}),
i.e., to compute $c_{0}$ and $c_{1}$ following Eqn.~(\ref{eq:ode2matrix-project-to-eigenvectors-how-to}).
The updated state at time $t=\Delta t$ is
\begin{align}
\mathbf{q}_{\parallel}\left(t=\Delta t\right)=\mathbf{q}_{\parallel}\left(t=0\right)+ & c_{0}\left[\begin{array}{c}
\omega_{p}\left[\cos\left(\omega_{p}\Delta t\right)-1\right]\\
\omega_{1}\sin\left(\omega_{p}\Delta t\right)\\
\omega_{2}\sin\left(\omega_{p}\Delta t\right)\\
\vdots\\
\omega_{N-1}\sin\left(\omega_{p}\Delta t\right)\\
\omega_{N}\sin\left(\omega_{p}\Delta t\right)
\end{array}\right]+c_{1}\left[\begin{array}{c}
-\omega_{p}\sin\left(\omega_{p}\Delta t\right)\\
\omega_{1}\left[\cos\left(\omega_{p}\Delta t\right)-1\right]\\
\omega_{2}\left[\cos\left(\omega_{p}\Delta t\right)-1\right]\\
\vdots\\
\omega_{N-1}\left[\cos\left(\omega_{p}\Delta t\right)-1\right]\\
\omega_{N}\left[\cos\left(\omega_{p}\Delta t\right)-1\right]
\end{array}\right].
\end{align}

\subsubsection{\label{subsec:exact-perp}Perpendicular Sub-System}

This section identifies the eigenvectors needed in the loop~\ref{lst:line:exact-for-par-perp} of Algorithm~(\ref{alg:exact-one-cell})
for the \emph{perpendicular} sub-system.
To get an elegant form of the sub-system, consider the
two perpendicular components of a vector as the real and imaginary
parts of a complex vector, i.e., $\tilde{\mathbf{E}}_{\perp}=\tilde{E}_{1}+i\tilde{E}_{2}$,
etc. Then we can write the perpendicular system as
\begin{equation}
\frac{\partial}{\partial t}\left[\begin{array}{c}
\tilde{\mathbf{E}}_{\perp}\\
\tilde{\mathbf{J}}_{1\perp}\\
\tilde{\mathbf{J}}_{2\perp}\\
\vdots\\
\tilde{\mathbf{J}}_{N-1,\perp}\\
\tilde{\mathbf{J}}_{N\perp}
\end{array}\right]=\mathbf{M}_{\perp}\cdot\left[\begin{array}{c}
\tilde{\mathbf{E}}_{\perp}\\
\tilde{\mathbf{J}}_{1\perp}\\
\tilde{\mathbf{J}}_{2\perp}\\
\vdots\\
\tilde{\mathbf{J}}_{N-1,\perp}\\
\tilde{\mathbf{J}}_{N\perp}
\end{array}\right],\label{eq:dE-dJ-perp-matrix-form}
\end{equation}
where the complex coefficient matrix is

\begin{equation}
\mathbf{M}_{\perp}=\left[\begin{array}{ccccc}
0 & -\omega_{1} & -\omega_{2} & \cdots & -\omega_{N}\\
\omega_{1} & -i\,\Omega_{1} & 0 & 0 & 0\\
\omega_{2} & 0 & -i\,\Omega_{2} & 0 & 0\\
\vdots & 0 & 0 & \ddots & 0\\
\omega_{N} & 0 & 0 & 0 & -i\,\Omega_{N}
\end{array}\right].\label{eq:perp-mat}
\end{equation}

Since $\mathbf{M}_{\perp}$ is skew-Hermitian, it has $N+1$ purely
imaginary eigenvalues $\lambda_j=i\omega_j$ that can be obtained (see
Sec.~\ref{subsec:appendix-exact-perp}) by solving

\begin{equation}
\omega-\sum_{s}\frac{\omega_{s}^{2}}{\omega+\Omega_{s}}=0.\label{eq:exact-perp-eigenvalue}
\end{equation}
Each eigenvalue has two real, orthogonal solution bases of different
polarizations (see Sec.~\ref{subsec:appendix-exact-perp}):
\begin{equation}
\left[\begin{array}{c}
\tilde{E}_{1}\\
\tilde{E}_{2}\\
\vdots\\
J_{s1}\\
\tilde{J}_{s2}\\
\vdots
\end{array}\right]=\left[\begin{array}{c}
-\sin\omega t\\
\cos\omega t\\
\vdots\\
\frac{\omega_{s}}{\omega+\Omega_{s}}\cos\omega t\\
\frac{\omega_{s}}{\omega+\Omega_{s}}\sin\omega t\\
\vdots
\end{array}\right]\text{ and }\left[\begin{array}{c}
\cos\omega t\\
\sin\omega t\\
\vdots\\
\frac{\omega_{s}}{\omega+\Omega_{s}}\sin\omega t\\
-\frac{\omega_{s}}{\omega+\Omega_{s}}\cos\omega t\\
\vdots
\end{array}\right].\label{eq:exact-perp-real-solutions}
\end{equation}
At $t=0$ they are 
\begin{equation}
\left[\begin{array}{c}
E_{1}\\
E_{2}\\
\vdots\\
u_{1s}\\
u_{2s}\\
\vdots
\end{array}\right]=\left[\begin{array}{c}
0\\
1\\
\vdots\\
\frac{\omega_{s}}{\omega+\Omega_{s}}\\
0\\
\vdots
\end{array}\right]\text{ and }\left[\begin{array}{c}
1\\
0\\
\vdots\\
0\\
-\frac{\omega_{s}}{\omega+\Omega_{s}}\\
\vdots
\end{array}\right].
\end{equation}
By the theory of Hermitian matrices, the eigenvectors of distinct
eigenvalues are orthogonal. Therefore we recover $2\times\left(N+1\right)$
orthogonal solution basis vectors.

Finally, the time-dependent solution to the original perpendicular sub-system
Eqn.~(\ref{eq:dE-dJ-perp-matrix-form}) can be obtained following
the ``project-and-evolve'' procedure outlined in Sec.~\ref{subsec:appendix-skew-hermitian-eigenvalue-problem}.

\subsection{Exact Solution to the Pressure Tensor Rotation}

The source term for the pressure tensor rotation, Eqn.~(\ref{eq:dPdt_source}),
can be written more explicitly as
\begin{equation}
\begin{split}\frac{d}{dt}\left[\begin{matrix}P_{xx}\\
P_{xy}\\
P_{xz}\\
P_{yy}\\
P_{yz}\\
P_{zz}
\end{matrix}\right]=\frac{q}{m}\left[\begin{array}{cccccc}
0 & 2B_{y} & -2B_{y} & 0 & 0 & 0\\
-B_{z} & 0 & B_{x} & B_{z} & -B_{y} & 0\\
B_{y} & -B_{x} & 0 & 0 & B_{z} & -B_{y}\\
0 & -2B_{z} & 0 & 0 & 2B_{x} & 0\\
0 & B_{y} & -B_{z} & -B_{x} & 0 & B_{x}\\
0 & 0 & 2B_{y} & 0 & -2B_{x} & 0
\end{array}\right]\left[\begin{matrix}P_{xx}\\
P_{xy}\\
P_{xz}\\
P_{yy}\\
P_{yz}\\
P_{zz}
\end{matrix}\right].\end{split}
\label{eq:dPdt-original}
\end{equation}
It can be solved analytically, too, and can be implemented following Algorithm ~(\ref{alg:exact-one-cell-pressure}).
\begin{algorithm}
\caption{Update source term for pressure tensor equation for species $s$ exactly in one cell}\label{alg:exact-one-cell-pressure}
\begin{algorithmic}[1]
\State Compute the rotation angle $\alpha\left(\Delta t\right)=-\Omega_{cs}\Delta t$ counter-clockwise around $\mathbf{B}$;  \label{lst:line:exact-pressure-angle}
\State Compute the rotation matrix $\mathbf{R}$ for $\alpha$; \label{lst:line:exact-pressure-matrix}
\State Rotate the pressure tensor $\mathbf{P}$ by applying the rotation matrix $\mathbf{R}$; \label{lst:line:exact-pressure-rotate}
\end{algorithmic}
\end{algorithm}
The rotation matrix $\mathbf{R}$ in line~\ref{lst:line:exact-pressure-matrix} of Algorithm ~(\ref{alg:exact-one-cell-pressure}) is
\begin{equation}
\mathbf{R}=\left[\begin{array}{ccc}
\cos\alpha+(1-\cos\alpha)b_{1}^{2}, & (1-\cos\alpha)b_{1}b_{2}-\sin\alpha\;b_{3}, & (1-\cos\alpha)b_{1}p_{3}+\sin\alpha\;b_{2}\\
(1-\cos\alpha)b_{2}b_{1}+\sin\alpha\;b_{3}, & \cos\alpha+(1-\cos\alpha)b_{2}^{2}, & (1-\cos\alpha)b_{2}p_{3}-\sin\alpha\;b_{1}\\
(1-\cos\alpha)b_{3}b_{1}-\sin\alpha\;b_{2}, & (1-\cos\alpha)b_{3}b_{2}+\sin\alpha\;b_{1}, & \cos\alpha+(1-\cos\alpha)b_{3}^{2}
\end{array}\right],\label{eq:exact-pressure-rotation-matrix}
\end{equation}
The pressure tensor rotation in line~\ref{lst:line:exact-pressure-rotate} can be performed as
\begin{equation}
\mathbf{P}\left(\Delta t\right)=\mathbf{R}\cdot\mathbf{P}\left(0\right)\cdot\mathbf{R}^{T}.\label{eq:exact-pressure-update}
\end{equation}
in the expanded form
\begin{align}
P_{mn}\left(t\right) & =\sum_{i}\sum_{j}P_{ij}\left(0\right)R_{mi}R_{nj}\nonumber \\
 & =P_{11}\left(0\right)R_{m1}R_{n1}+P_{21}\left(0\right)R_{m2}R_{n1}+P_{31}\left(0\right)R_{m3}R_{n1}\nonumber \\
 & +P_{12}\left(0\right)R_{m1}R_{n3}+P_{22}\left(0\right)R_{m2}R_{n2}+P_{32}\left(0\right)R_{m3}R_{n2}\nonumber \\
 & +P_{13}\left(0\right)R_{m1}R_{n3}+P_{23}\left(0\right)R_{m2}R_{n3}+P_{33}\left(0\right)R_{m3}R_{n3}.
\end{align}

\section{\label{sec:implicit}A Locally Implicit Scheme}

The exact source solutions obtained in the Section.~\ref{sec:exact}
are relatively expensive to compute. More importantly, it is difficult to incorporate
additional source terms as they might change the fundamental structure
of the linear system. In this section, we develop a \emph{locally
implicit} scheme to update the source term equations more efficiently
using a centered discretization in time. Using this scheme, it is also straightforward
to include additional source terms that depend on local quantities only,
e.g., collisions and ionization. It is worthwhile to mention that
schemes using implicit\textendash explicit (IMEX) timestepping to
treat the source terms are described in \citep{Kumar2012,Abgrall2014,Balsara2016a}.
A more recent work coupled a biased implicit treatment of the source term
in with the hyperbolic update\citep{Huang2019}. However the scheme
presented here is considerably simpler as we work with the non-conservative
form of the equations \emph{just for the source updates}. In any case,
the scheme in \citep{Kumar2012} is implicitly contained in earlier
two-fluid papers \citep{sumlak_2003,Hakim2006}, which essentially
only performed a single (or few) iteration(s) of the implicit scheme
in \citep{Kumar2012}. 

\subsection{The Scheme}

For numerical stability, it is intuitive to apply the backward Euler
method,
\begin{equation}
\begin{cases}
\mathbf{J}_{s}^{n+1} & =\mathbf{J}_{s}^{n}+\frac{\Delta t}{2}\left(\omega_{s}^{2}\varepsilon_{0}\mathbf{E}^{n+1}+\mathbf{J}_{s}^{n+1}\times\mathbf{\Omega}_{s}\right)\\
\mathbf{E}^{n+1} & =\mathbf{E}^{n}-\frac{\Delta t}{2\varepsilon_{0}}\sum_{s}\mathbf{\bar{J}}_{s}.
\end{cases}\label{eq:dJdt_dEdt_implicit-backward-Euler}
\end{equation}
which was adopted by \citep{Abgrall2014}. However, this is a first
order method and damps the oscillatory solutions. A slight modification
improves both stability and energy conservation. For convenience,
we introduce time-centered quantities
\begin{equation}
\mathbf{\bar{J}}_{s}\equiv(\mathbf{J}_{s}^{n+1}+\mathbf{J}_{s}^{n})/2\quad{\rm and}\quad\mathbf{\bar{E}}\equiv(\mathbf{E}^{n+1}+\mathbf{E}^{n})/2
\end{equation}
 The first part of our scheme is given by 
\begin{equation}
\begin{cases}
\mathbf{\bar{J}}_{s} & =\mathbf{J}_{s}^{n}+\frac{\Delta t}{2}\left(\omega_{s}^{2}\varepsilon_{0}\mathbf{\bar{E}}+\mathbf{\bar{J}}_{s}\times\mathbf{\Omega}_{s}\right)\\
\mathbf{\bar{E}} & =\mathbf{E}^{n}-\frac{\Delta t}{2\varepsilon_{0}}\sum_{s}\mathbf{\bar{J}}_{s}
\end{cases}\label{eq:dJdt_dEdt_implicit}
\end{equation}
and can be rearranged into the form
\begin{equation}
\begin{cases}
\mathbf{\bar{J}}_{s}-\frac{\Delta t}{2}\left(\omega_{s}^{2}\varepsilon_{0}\mathbf{\bar{E}}+\mathbf{\bar{J}}_{s}\times\mathbf{\Omega}_{s}\right) & =\mathbf{J}_{s}^{n}\\
\mathbf{\bar{E}}+\frac{\Delta t}{2\varepsilon_{0}}\sum_{s}\mathbf{\bar{J}}_{s} & =\mathbf{E}^{n}
\end{cases}\Leftrightarrow\mathbf{M}_{l.h.s.}\left[\begin{array}{c}
\mathbf{\bar{J}}_{s}\\
\mathbf{\bar{E}}
\end{array}\right]=\left[\begin{array}{c}
\mathbf{\mathbf{J}}_{s}\\
\mathbf{\mathbf{E}}
\end{array}\right]\label{eq:dJdt_dEdt_implicit-matrix-form}
\end{equation}
with $\mathbf{M}_{l.h.s.}$ being the $(3S+3)\times(3S+3)$ constant coefficient matrix.

Eqn.~(\ref{eq:dJdt_dEdt_implicit-matrix-form}) is a system of linear, constant-coefficient ODEs for the $3S+3$
unknowns $\mathbf{\bar{J}}_{s}$ and $\mathbf{\bar{E}}$ and can be
solved with any linear algebra routine to get
\begin{equation}
\left[\begin{array}{c}
\mathbf{\bar{J}}_{s}\\
\mathbf{\bar{E}}
\end{array}\right]=\mathbf{M}_{l.h.s.}^{-1}\left[\begin{array}{c}
\mathbf{\mathbf{J}}_{s}\\
\mathbf{\mathbf{E}}
\end{array}\right].
\end{equation}
The final updated currents and electric
field can then be determined by
\begin{equation}
\mathbf{E}^{n+1}=2\mathbf{\bar{E}}-\mathbf{E}^{n}\quad{\rm and}\quad\mathbf{J}_{s}^{n+1}=2\mathbf{\bar{J}}_{s}-\mathbf{J}_{s}^{n}.
\end{equation}
We call this scheme the locally implicit scheme as it involves only
data in a single cell and requires the inversion of only a $(3S+3)^2$ matrix.
No global matrix inversion coupling all
cells in the domain is required. If a DG scheme is used, then the
source update needs to be computed at each node of the selected finite-element
node, or, projected onto the selected modal basis, if using those.
In Section ~\ref{subsec:appendix-implicit-direct}, we show that
it is also possible to write down the resultant formulae for the complete
linear algebra calculations and skip the ``null'' calculations (e.g., zeros
multiplied by zeros) for significant speedup.

\subsection{Accuracy and Stability}

Indeed, the locally implicit scheme is an implicit midpoint method,
giving an error of order $O\left(\left(\Delta t\right)^{2}\right)$.
The stability of the algorithm can be studied by a von Neumann analysis.
We introduce a time-dependence of $e^{-i\omega t}$, where $\omega$
is the (possibly complex) numerical frequency. For plasma oscillations
the source update has the numerical dispersion relation (see Section.~\ref{subsec:stability})
\begin{equation}
4/\Delta t^{2}\tan^{2}(\omega\Delta t/2)=\omega_{p}^{2}
\end{equation}
showing that the time-step is not restricted by plasma frequency.
In a similar way, we can show that the time step is not restricted
by cyclotron frequency, either.

\subsection{Conservation Properties}

To ensure that the number density and pressure remain positive, we
first observe that the source update Eqns.\,(\ref{eq:dJdt_dEdt_implicit-with-collision})
do not modify either of these quantities. Hence, positivity violations
can only occur in the homogeneous updates of the fluid quantities. 

The multifluid system, in the absence of dissipation and with appropriate
boundary conditions\footnote{Energy conservation for \emph{homogeneous} fluid equations is exact
for periodic boundaries. However, for wall boundary small energy errors
arise due to diffusive terms in the numerical fluxes used. One can
always use a central flux at walls, but this complicates the scheme,
and is not always worth the effort in practice.}, conserves the total energy, i.e $d\mathcal{E}/dt=0$, where 
\begin{align}
\mathcal{E}=\int\left[\sum_{s}\left(\frac{1}{2}m_{s}\mathbf{u}_{s}^{2}+\frac{3}{2}p_{s}\right)+\frac{\epsilon_{0}}{2}\mathbf{E}^{2}+\frac{1}{2\mu_{0}}\mathbf{B}^{2}\right]d\mathbf{x}
\end{align}
and the integration is taken over the whole domain. The source update,
Eqns.\,(\ref{eq:dJdt_dEdt_implicit})
as can be shown easily, conserve the discrete form of Eqn.~(\ref{eq:src-energy}).
This can be further understood as the merit of the implicit midpoint
method that it preserves the magnitude of any oscillatory systems.
Also, solving the homogeneous fluid equation in conservation law form
conserves the fluid energies. Hence, the conservation of the total
discrete energy, including electromagnetic energy, depends on the
scheme selected to solve Maxwell equations. In general, upwind finite-volume
schemes will \emph{not} conserve the EM energy, but \emph{decay} it.
Hence, unless an energy conserving finite-volume/difference scheme
is used to update the EM fields, the total energy is not conserved
by the discrete scheme. Even when using an upwind scheme, lack of
energy conservation is not always a problem, however, as the energy
conservation error scales as the order of the scheme, and hence can
be controlled by using a finer mesh or a higher order method.

\subsection{Including Collisions}

Following \citep{Smithe2007a}, we may incorporate frictional collisions
by slightly modifying the current source term as
\begin{equation}
\left(\frac{\partial}{\partial t}+\nu_{s}\right)\mathbf{J}_{s}=\omega_{s}^{2}\varepsilon_{0}\mathbf{E}+\mathbf{J}_{s}\times\boldsymbol{\Omega}_{s}\label{eq:dJdt_dEdt_source-with-collision}
\end{equation}
where $\nu_{s}$ is a constant collision frequency. The additional
collision term converts the solution from purely oscillatory to damped
oscillations. The corresponding locally implicit scheme is written as
\begin{equation}
\begin{cases}
\mathbf{\bar{J}}_{s} & =\mathbf{J}_{s}^{n}+\frac{\Delta t}{2}\left(\omega_{s}^{2}\varepsilon_{0}\mathbf{\bar{E}}+\mathbf{\bar{J}}_{s}\times\mathbf{\Omega}_{s}-\nu_{s}\bar{\mathbf{J}}_{s}\right)\\
\mathbf{\bar{E}} & =\mathbf{E}^{n}-\frac{\Delta t}{2\varepsilon_{0}}\sum_{s}\mathbf{\bar{J}}_{s}.
\end{cases}\label{eq:dJdt_dEdt_implicit-with-collision}
\end{equation}
This is still a constant coefficient linear system, thus can also
be solved using any linear algebra routine. In a similar manner, we
may include additional source terms, e.g., ionization, gravity, chemical
production, as long as
the terms involves only local quantities (i.e., no gradient calculations
etc.).

\section{\label{sec:benchmark}Benchmark Problems}

In this section, we present a few benchmark simulations to illustrate the properties of the locally implicit scheme. In all but the first test, we use a dimensionally split version of the FV discretization
described in \citep{Hakim2006}. Note that the purpose of
the paper is not to study the \emph{detailed physics} of these problems,
but to simply show that the algorithm presented above is efficient and stable where
the explicit algorithm would be unstable, and produces qualitatively
correct results.

\subsection{Plasma Oscillation}

\begin{figure}
\begin{centering}
\includegraphics[width=0.98\textwidth]{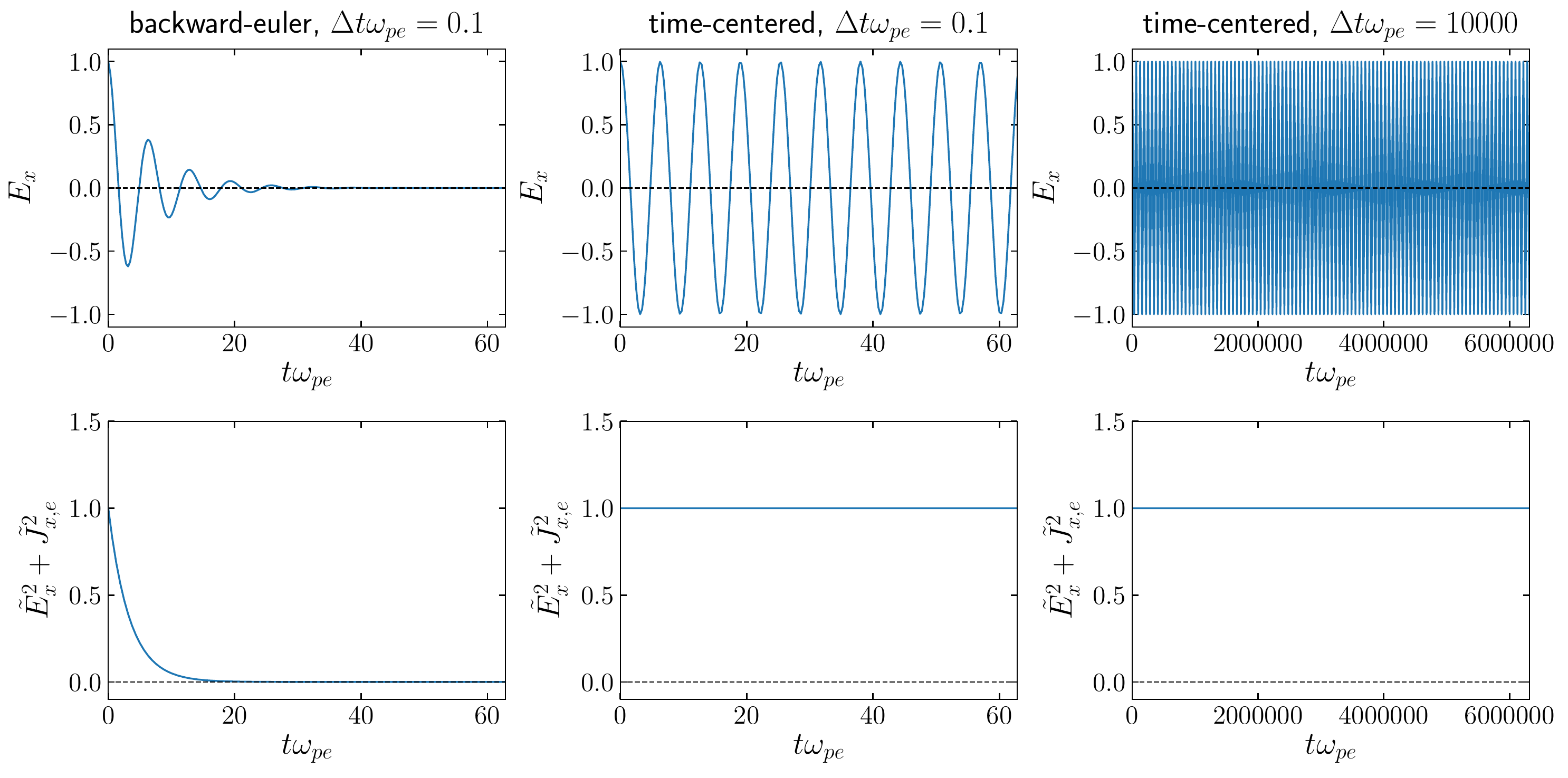}
\par\end{centering}
\caption{\label{fig:plasma-oscillation-test}Electron plasma oscillation using
the first-order backward Euler scheme (left column) and locally implicit time-centered algorithm (middle and right columns).
\textit{Upper row}: Temporal evolution of $E_{x}$. \textit{Lower row}:
Temporal evolution of total normalized energy $\tilde{E}_{x}^2+\tilde{J}_{x,e}^2$. The time-centered simulations
are fully oscillatory without any damping or instability.}

\end{figure}

As our first benchmark, we consider the plasma oscillation of one species locally, i.e., within a cell. In other word, we disregard the spatial integration Eqn.~(\ref{eq:homosys}) and integrates only Eqn.~(\ref{eq:dJdt_dEdt_source}). The purpose is to verify the conservation property of the time-centered scheme. The initial condition consists of a uniform, stationary plasma and vanishing magnetic field. Perturbation is imposed on $E_{x}$. Therefore the initial condition is an eigenvector of $\omega=\omega_{p}$ (see the exact solution Eqn.~(\ref{eq:exact-para-eigenvectors-at-t})). The ions are assumed to be immobile.
We performed three simulations as presented in the three columns in Figure.~(\ref{fig:plasma-oscillation-test}). The upper and lower rows are the time evolution of $E_x$ and the phase-diagram for the normalized electric field $\tilde E_x$ and current $\tilde J_{x,e}$ (see Eqn.~(\ref{eq:Enorm-Jnorm})), respectively. Ideally, the system should oscillate at the plasma frequency $\omega_{pe}$ and the total normalized energy $\tilde{E}_x^2+\tilde{J}_{x,e}^2$ should remain constant following energy conservation.

The simulation in the left column uses the first-order backward-Euler implicit algorithm. Even with a small time step $\Delta t=0.1/\omega_{pe}$, the solution is quickly damped.
In comparison, the middle-column simulation that uses the time-centered implicit algorithm is able to model the oscillation at the correct frequency and does not suffer from any damping.
The conservation property of the time-centered algorithm can be further confirmed from the right-column run that uses a large time step $\Delta t=10000/\omega_{pe}$ with the time-centered algorithm (see the lower panel). At such a large time step, however, the plasma oscillation are aliased by slower oscillations that are supported by the time step size.

\subsection{Plasma Wave-Beach}

\begin{figure}
\centering{} \includegraphics[width=0.6\textwidth]{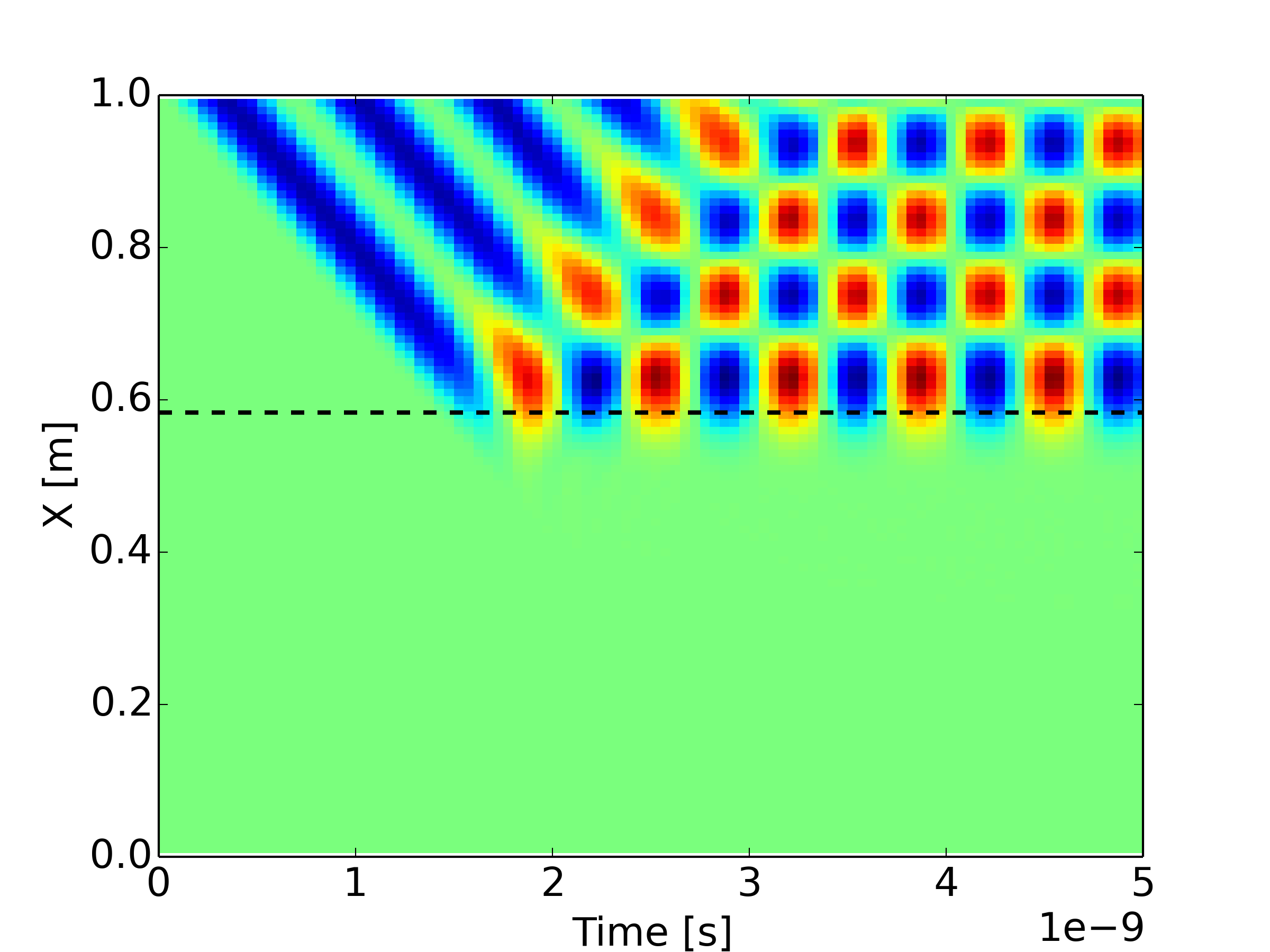}\caption{\label{fig:bench-plasma-beach}Propagation of an electromagnetic (EM)
wave into a ``plasma wave-beach''. Shown are contours of the $y$-component
of the electric field, time increasing from left to right. The EM
wave is launched by applying a current ($J_{y}$) at $x=1$, with
frequency $\omega\delta t=\pi/10$, with $\delta t=1/100c$. The wave
suffers a cutoff (black dashed line), reflecting back into the low-density
region, creating a standing wave pattern late in time. The time-step
is $12.5\times$ larger than for a fully explicit scheme, which would
have a restriction $\omega_{e}\Delta t<2$.}
 
\end{figure}

We now show a few more practical problems that couple the homogeneous
part and the source term part. The first is a ``plasma wave-beach''
problem, in which power is propagated in a $1$~m long hydrogen plasma
of increasing density, such that the wave suffers a cutoff at $x=0.58$~m.
Letting $\delta t=1/100c$, the plasma profile is $\omega_{e}(x)\delta t=(1-x)^{5}$.
For the $100$~cell simulation shown in Figure.\,(\ref{fig:bench-plasma-beach}),
hence, the time-step, restricted only by the CFL condition, is $\omega_{e}\Delta t=25$,
$12.5$ times larger than would be allowed by a fully explicit scheme,
which has a restriction $\omega_{e}\Delta t<2$.

\subsection{Magnetic Reconnection in a Harris Current Sheet}

The second simulation is of magnetic reconnection in a Harris current
sheet. For this, the standard GEM reconnection challenge parameters
are used \citep{birn_2001a}, with an initial equilibrium magnetic
field $B_{x}(y)=B_{0}\tanh(y/L)$, supported by a out-of-plane current
sheet with both electrons and ions carrying current. The simulation
parameters are 
\begin{align}
\frac{\lambda}{d_{i}}=0.5,\ \frac{m_{i}}{m_{e}}=100,\ \frac{T_{i}}{T_{e}}=5,\ \frac{n_{b}}{n_{0}}=0.2,\ \frac{V_{A0}}{c}=0.05
\end{align}
where $d_{i}=c/\omega_{pi}$ is the ion inertial length, $T_{i}$
and $T_{e}$ are the ion and electron temperatures, and $v_{A0}=B_{0}/\sqrt{\mu_{0}n_{0}m_{i}}$
is the Alfv\'en velocity. The plasma beta is unity. A grid of $64\times32$
cells was used, with a CFL number of $0.9$, resulting in a time-step
of $\omega_{pe}\Delta t\approx3.6$. The cell spacing is about $27\times$
larger than the Debye length. Even on this coarse mesh, with the plasma
frequency unresolved and the Debye length severely under-resolved,
the algorithm is stable and produces results qualitatively similar
to higher resolution results published in \citep{Hakim2006,Loverich:2010ea},
clearly showing the reconnected current sheet structure, as well
as the quadrupolar out-of-plane magnetic field formed due to Hall currents.
More thorough studies of magnetic reconnection in the context of Earth's
magnetosphere, etc. can be found in \citep{TenBarge2019} and \citep{Wang2015}.

\begin{figure}
\begin{centering}
\includegraphics[width=0.75\textwidth]{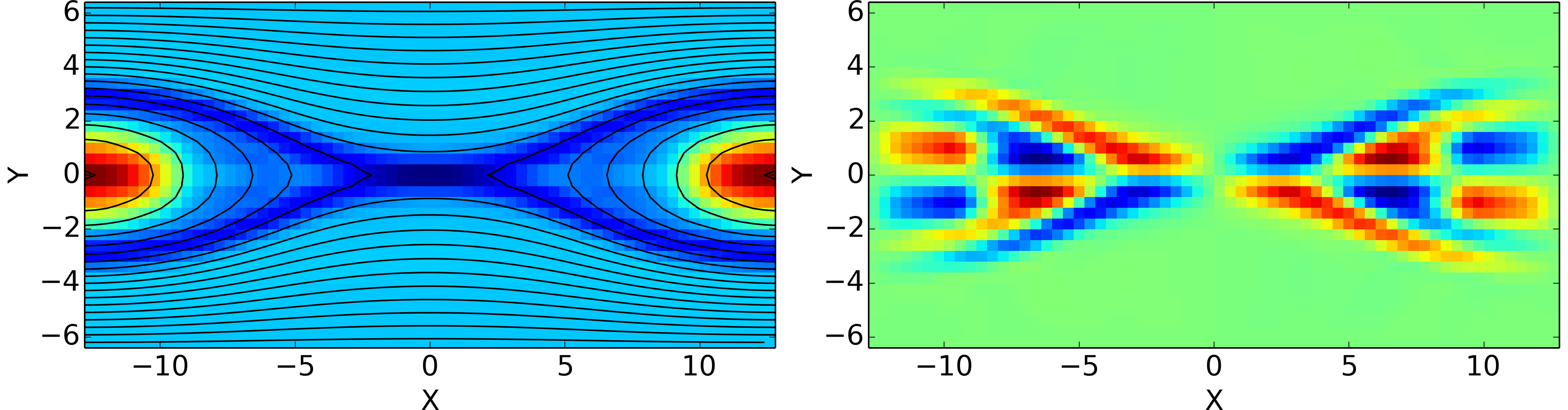}
\par\end{centering}
\caption{Out-of-plane electron current and contours of in-plane magnetic field
(upper), and out-of-plane magnetic field (lower) from a GEM reconnection
simulation. Quadrupole structure of the Hall magnetic field is clearly
visible. A $64\times32$ grid was used, for which the $\omega_{e}\Delta t\approx3.6$
and the cell spacing is about $27\times$ larger than the Debye length.
Even on this coarse mesh, with the plasma frequency unresolved and
the Debye length severely under-resolved, the algorithm is stable
and produces results qualitatively similar to previously published
five-moment results\citep{Hakim2006,Loverich:2010ea}.}
\label{fig:gem-recon} 
\end{figure}

\subsection{Orszag-Tang Vortex}

\begin{figure}
\includegraphics[width=1\columnwidth]{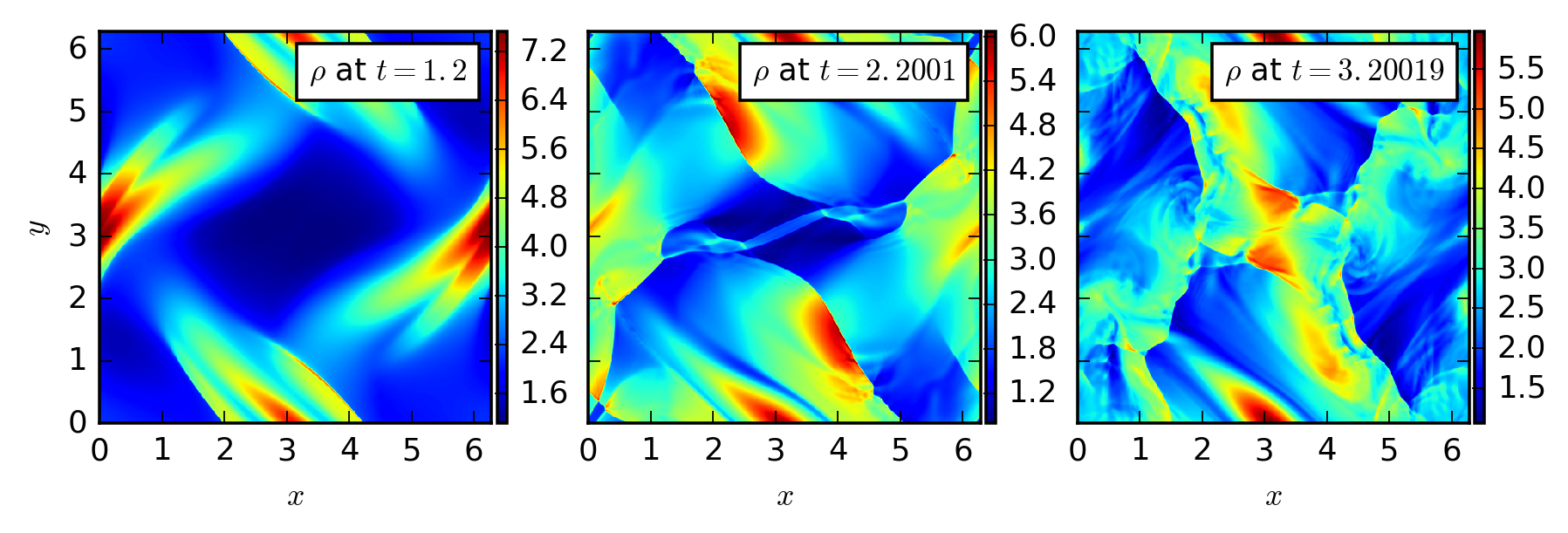}

\caption{\label{fig:ot}Total mass densities at different stages in the five-moment
Orszag-Tang vortex test.}
\end{figure}

The fourth test is five-moment simulation of he Orszag-Tang vortex \citep{Orszag1979},
a 2D problem extensively used to benchmark and compare numerical codes
\citep{Toth2000b,Stone2008,Dudson2009}. We use a $\left[0,2\pi\right]\times\left[0,2\pi\right]$
periodic domain on a $512\times512$ grid. The initial condition consists
of uniform total mass density $\rho=25/9$, uniform total pressure
$p=5/3$, in-plane flow vortex $v_{x}=-\sin y$, $v_{y}=\sin x$,
and magnetic field vortex $B_{x}=-\sin y$, $B_{y}=\sin\left(2x\right)$.
The ion charge/mass ratio is so that the ion inertia lengths based
on initial background density is $d_{i}=\sqrt{\rho_{i}q_{i}^{2}/\varepsilon_{0}m_{i}^{2}}=2\pi/25$.
Other parameters include $\gamma=5/3$, $\mu_{0}=1$, $c=20$, $m_{i}/m_{e}=25$,
and $p_{i}/p_{e}=1$. As shown in Figure.\,(\ref{fig:ot}), the formation
of shocks is clearly captured, and the strong shock-shock interactions
produce rather dynamic turbulence.

\subsection{Solar Wind-Magnetosphere Interaction at the Earth}

Finally, we present results of 3D, large scale simulation of the Earth's
magnetosphere under the impact of solar wind plasmas transported from
the sun. The system is dominated by magnetic reconnection. At the
dayside, this happens between the Earth's dipolar magnetic field and
the southward interplanetary magnetic field. At the nightside, it
is between the highly stretched dipolar field. Later, we will demonstrate
the tail reconnection as an example.

The simulation domain has the Earth at the origin, the $x$-axis is
along the Sun-Earth line, and the $z$-direction is northward. The
domain spans $\left[-17,\,63\right]$ along $x$ in units of the Earth's
radii, and $\left[-81,\,81\right]$ in the other two directions. A
stretched nonuniform grid of total size $1600\times2200\times2200$
is used to achieve high resolution near the Earth and near the day-
and night-side reconnection sites. During the simulation, the upstream
conditions at $x=-17$ are fixed solar wind parameters. All other
boundaries are open and perturbations are allowed to exit the domain.
The simulation lasted 3600$s$ in physical time, and took about 2
million core hours to finish on Pleiades, a petascale supercomputer
housed at the NASA Advanced Supercomputing (NAS) facility.

\begin{figure}
\begin{centering}
\includegraphics[width=0.9\textwidth]{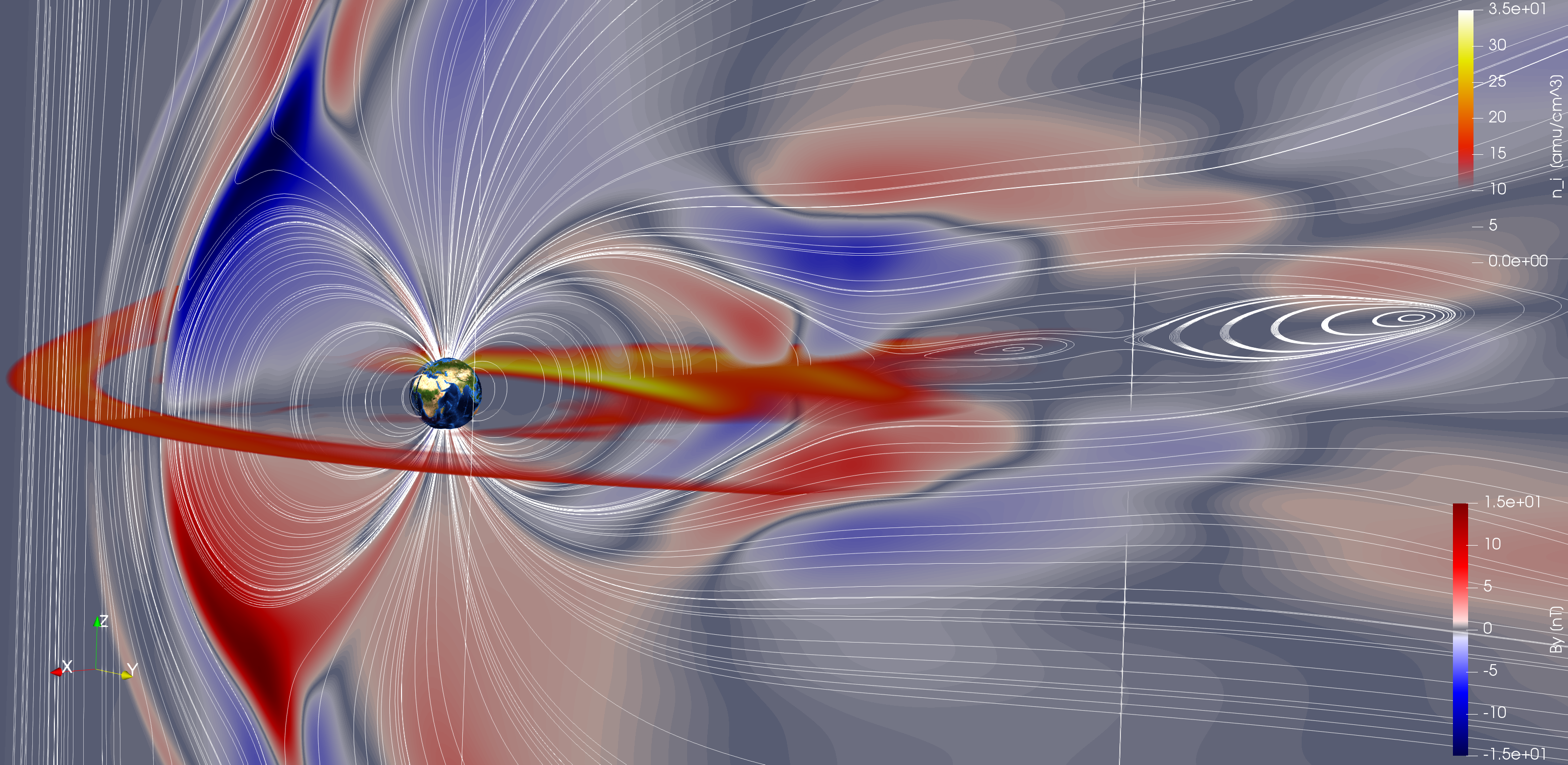}
\par\end{centering}
\caption{\label{fig:bench-earth}Perspective view of the Earth's magnetosphere from a Ten-Moment simulation. The solar wind impinges 
from the left side and carries southward interplanetary magnetic 
field.
The white streamlines are magnetic field lines in the meridional plane ($xz$-plane, which is approximately the paper plane), indicating the thinning of a current sheet on the night-side, along with two plasmoids.
The blue-red contours in the $xz$-plane represent the $y$ component of the magnetic field
due to the Hall term intrinsic to the model.
The yellow-scale contours in the $xy$-plane represent ion number density. Its sharp
boundary at the dayside (left side) marks the bow shock due to the supersonic and superAlfv\'enic solar wind flow.
}

\end{figure}

Figure~(\ref{fig:bench-earth}) shows a perspective view snapshot.
The yellow-white-coded contours represent the
ion number density in the equatorial plane. Their dayside boundaries
mark the sharp shock due to the supersonic and superAlfv\'enic inflow.
The blue-red-coded contours are the non-vanishing $B_{y}$ in the $xz$
plane due to the Hall effect contained in the multifluid model. The
white streamlines are magnetic field lines in the same plane. 
The snapshot clearly shows the formation 
of plasmoids, an coherent structure containing isolated regions of
magnetic fluxes. The relaxation of field lines after ejection and
the birth of a new plasmoid are depicted in the last two frames, indicating
a repeating life-cycle of the system. Planetary/moon magnetospheric
physics are often quite complicated. For more in-depth investigations,
interested readers may refer to \citep{Wang2018jgr,Dong2019} for
applications of this model to other magnetosphere systems.

\section{\label{sec:conclusions}Conclusions}

Multifluid-Maxwell models have been rapidly gaining interest recently
in laboratory, space, and astrophysical plasma phyiscs modeling\citep{Wang2015,Ng2015,Huang2019,Amano2015,Balsara2016a,Hirabayashi2016,Zenitani2009,Allmann-Rahn2018a,Lautenbach2018,Joncquieres,laguna2017effect,alvarez2018versatile,laguna2018fully}.
In this paper, we have developed a locally implicit schemes to integrate
the source terms for such systems. Numerical restrictions due to kinetic
scales like plasma and cyclotron frequencies are eliminated. The
calculation involves only local (one-cell) inversion of a $3S+3$
matrix, where $S$ is the number of species. Direct formulae that
avoid null calculations in full matrix operations are derived that
can be used to speed up the algorithm. The stability of the source
term solver alone and the complete coupled system is validated using
a few benchmark problems, ranging from small, idealized problems to
large, complex systems. For completeness, we also derived the
exact solutions to the source term equations, which is often more
expensive to compute but nevertheless can be implemented as a base
benchmark solver.

Eliminating the restrictions due to kinetic scales is crucial for
the practical application of the multifluid-Maxwell model. The coupled
system is still constrained by the CFL condition due to speed of light,
but this is often a less demanding requirement. The locally implicit
algorithm enables us to model large, multi-scale systems by stepping
over smallest scale physics that might not be crucial for the global
picture. In fact, we have successfully applied our model to plasmas
in a vast range of problems\citep{Wang2015,Wang2018jgr,Ng2015,Dong2019,TenBarge2019,Ng2017b,ng2018pop,Ng2019,Cagas2017a,Ng2018a}.
The capability to retain finite order kinetic physics (Hall term,
electron inertia, pressure non-gyrotropy etc.) of the coupled multifluid-Maxwell
model is desirable for the study of many problems, e.g., magnetic
reconnection and turbulence. It is of course also convenient to be able to incorporate
an arbitrary number of plasma species, particularly electron physics
and multi-ion effects.

\section*{Acknowledgements}

The authors thank Dr.~Peifeng Fan for useful discussions.
This work was supported by Air Force Office of Scientific Research under Grant No. FA9550-15-1-0193, DOE grant No. DEAC02-09CH11466, NSF Grant Nos. AGS0962698 and AGS-1338944, NASA Grants Nos. NNH13AW51I, 80NSSC19K0621, 80NSSC18K0288 and  NNX13AK31G.
Resources supporting this work were provided by the NASA High-End Computing
(HEC) Program through the NASA Advanced Supercomputing (NAS) Division at Ames Research Center,
the National Energy Research Scientific Computing Center, a DOE Office of Science User Facility
supported by the Office of Science of the U.S. Department of Energy under Contract No. DE-AC02-05CH11231,
Cheyenne (doi:10.5065/D6RX99HX) provided by NCAR's CISL, sponsored by NSF, and Trillian,
a Cray XE6m-200 supercomputer at the UNH supported by the NSF MRI program under Grant No. PHY-1229408.

\appendix
\renewcommand*{\thesection}{\Alph{section}}

\section{\label{subsec:appendix-skew-hermitian}Useful Notes on the Matrix
Form of the Linear System of Equations}

\subsection{Notes on Skew-Symmetric and Skew-Hermitian Matrices}

A square matrix $\mathbf{M}$ is skew-Hermitian if and only if its
conjugate transpose is its negative, 
\begin{equation}
M_{ij}^{*}=-M_{ji}.
\end{equation}
Here the superscript $^{*}$ represents the conjugate operation.

The following properties of a skew-Hermitian matrix $\mathbf{M}$
are useful in our derivation:
\begin{itemize}
\item $\mathbf{M}$ is diagonalizable and all of its eigenvalues are either
purely imaginary or zero.
\item Eigenvectors of distinct eigenvalues of $\mathbf{M}$ are orthogonal
to each other.
\item If $\mathbf{M}$ is also real, then its nonzero eigenvalues come in
pairs $\pm\lambda$.
\begin{itemize}
\item Now if $\mathbf{v}$ is an eigenvector of $\mathbf{M}$ with eigenvalue
$\lambda=i\omega$ then $\mathbf{v}^{*}$ is an eigenvector for eigenvalue
$-i\omega$.
\end{itemize}
\item $i\mathbf{M}$ is Hermitian.
\item If $\mathbf{v}$ is an eigenvector of $\mathbf{M}$ with eigenvalue
$\lambda=i\omega$, it is an eigenvalue of $i\mathbf{M}$ with eigenvalue
$\omega$.
\end{itemize}

\subsection{\label{subsec:appendix-skew-hermitian-eigenvalue-problem}Equivalent
Eigenvalue Problem}

Note that solving the constant-coefficient linear ODE system 
\begin{equation}
\frac{\partial\mathbf{q}}{\partial t}=\mathbf{M}\cdot\mathbf{q}.\label{eq:ode2matrix-ode-general}
\end{equation}
where $\mathbf{M}$ is a skew-Hermitian matrix, is equivalent to solving
the time-independent eigenvalue problem
\begin{equation}
\mathbf{M}\cdot\mathbf{v}_{\ell}=\mathbf{v}_{\ell}\lambda_{\ell}
\end{equation}
Here $\lambda_{\ell}$ and $\mathbf{v}_{\ell}$ are the $\ell$th
eigenvalue and eigenvector of $\mathbf{M}$.

To solve the initial value problem, we need to first compute the eigencoefficients
$c_{\ell}$ by projecting the initial condition onto the eigenvectors,
\begin{equation}
\mathbf{q}\left(t=0\right)=\sum_{\ell}c_{\ell}\mathbf{v}_{\ell}.\label{eq:ode2matrix-project-initial-to-eigenvectors}
\end{equation}
The temporal evolution then follows
\begin{equation}
\mathbf{q}\left(t\right)=\sum_{\ell}c_{\ell}\mathbf{v}_{\ell}\exp\left(\lambda_{\ell}t\right).\label{eq:ode2matrix-temporal-solution}
\end{equation}
Since $\mathbf{M}$ is skew-Hermitian, $\lambda_{\ell}$ must be purely
imaginary or zero. Therefore the solution above is oscillatory or
constant in time.

In addition, due to the theory of skew-Hermitian matrices, their eigenvectors
for distinct eigenvalues must be orthogonal. Therefore we can compute
the coefficients for nondegenerate eigenvectors by 
\begin{equation}
c_{\ell}=\mathbf{q}\left(t=0\right)\cdot\mathbf{v}_{\ell}.\label{eq:ode2matrix-project-to-eigenvectors-how-to}
\end{equation}
Note that the degenerate eigenvectors might not be orthogonal, though.

\subsection{\label{subsec:appendix-skew-hermitian-real-solutions}Constructing
Real Solutions}

The solution Eqn.~(\ref{eq:ode2matrix-temporal-solution}) can be
complex. Thus we need to construct real solutions from them to represent
real physical values. To do this, we first consider a complex eigenvector
$\mathbf{v}=\mathbf{a}+i\,\mathbf{b}$ of an eigenvalue $\lambda=i\omega$,
the corresponding time-dependent, complex solution to Eqn.~(\ref{eq:ode2matrix-ode-general})
is
\begin{align}
\mathbf{v}\exp\left(i\omega t\right) & =\left(\mathbf{a}+i\mathbf{b}\right)\left(\cos\omega t+i\sin\omega t\right)\nonumber \\
 & =\mathbf{a}\cos\omega t-\mathbf{b}\sin\omega t+i\left(\mathbf{a}\sin\omega t+\mathbf{b}\cos\omega t\right).\label{eq:ode2matrix-complex-solution-v}
\end{align}
Note that $-i\mathbf{v}$ is also an eigenvector with eigenvalue $i\omega$
and has the time-dependent solution
\begin{align}
\mathbf{v}\exp\left(i\omega t\right) & =\left(\mathbf{b}-i\mathbf{a}\right)\left(\cos\omega t+i\sin\omega t\right)\nonumber \\
 & =\mathbf{a}\sin\omega t+\mathbf{b}\cos\omega t-i\left(\mathbf{a}\cos\omega t-\mathbf{b}\sin\omega t\right).\label{eq:ode2matrix-complex-solution-minus-iv}
\end{align}
Thus we identify two independent real solution bases
\begin{equation}
\mathbf{a}\cos\omega t-\mathbf{b}\sin\omega t\text{ and }\mathbf{a}\sin\omega t+\mathbf{b}\cos\omega t\label{eq:ode2matrix-real-solution-bases}
\end{equation}
which are the real and imaginary parts of Eqn.~(\ref{eq:ode2matrix-complex-solution-v}).
In other words, to construct the real solutions, we only need to take
the real and imaginary parts of an complex eigenvector Eqn.~(\ref{eq:ode2matrix-complex-solution-v}).

\section{\label{sec:appendix-exact}Deriving the Exact Solutions}

\subsection{Identifying the renormalizations}

The normalization relation Eq.~(\ref{eq:Js0-E0-relation}) is identified
by observing that Eq.~(\ref{eq:dE-dJ-matrix}) in the renormalized
variables are
\begin{equation}
\frac{\partial}{\partial t}\left[\begin{array}{c}
\tilde{\mathbf{E}}\\
\tilde{\mathbf{J}}_{1}\\
\tilde{\mathbf{J}}_{2}\\
\vdots\\
\tilde{\mathbf{J}}_{N}
\end{array}\right]=\left[\begin{array}{ccccc}
0 & -\frac{J_{1,0}}{\varepsilon_{0}E_{0}} & -\frac{J_{2,0}}{\varepsilon_{0}E_{0}} & \cdots & -\frac{J_{N0}}{\varepsilon_{0}E_{0}}\\
\omega_{1}^{2}\frac{\varepsilon_{0}E_{0}}{J_{1,0}} & -\boldsymbol{\Omega}_{1}\times\boldsymbol{\mathcal{I}} & 0 & \cdots & 0\\
\omega_{2}^{2}\frac{\varepsilon_{0}E_{0}}{J_{2,0}} & 0 & -\boldsymbol{\Omega}_{2}\times\boldsymbol{\mathcal{I}} & \cdots & 0\\
\vdots & \vdots & \vdots & \ddots & \vdots\\
\omega_{N}^{2}\frac{\varepsilon_{0}E_{0}}{J_{N0}} & 0 & 0 & \cdots & -\boldsymbol{\Omega}_{N}\times\boldsymbol{\mathcal{I}}
\end{array}\right]\left[\begin{array}{c}
\varepsilon_{0}\tilde{\mathbf{E}}\\
\tilde{\mathbf{J}}_{1}\\
\tilde{\mathbf{J}}_{2}\\
\vdots\\
\tilde{\mathbf{J}}_{N}
\end{array}\right]\label{eq:dE-dJ-matrix-normalized-nominal}
\end{equation}
To anti-symmetrize the coefficient matrix, we require
\begin{equation}
-\omega_{s}^{2}\frac{\varepsilon_{0}E_{0}}{J_{s0}}=-\frac{J_{s0}}{\varepsilon_{0}E_{0}},
\end{equation}
hence Eq.~(\ref{eq:Js0-E0-relation}). Note that Eq.~(\ref{eq:Js0-E0-relation})
does not give specific values for $J_{s0}$ and $E_{0}$ but only
their relation. One may choose a nominal normalization, say, $E_{0}\equiv1$
and accordingly, $J_{s0}\equiv\varepsilon_{0}\omega_{ps}$.

\subsection{Solving the Parallel Sub-System}

Solving the eigenvalue problem of the parallel sub-system Eq.~(\ref{eq:dE-dJ-para-matrix-form})
\begin{equation}
\det\left|\mathbf{M}_{\parallel}-\lambda\right|=0
\end{equation}
gives three distinct eigenvalues
\begin{equation}
\left[-i\,\omega_{p},i\,\omega_{p},0\right]
\end{equation}
and their multiplicities
\begin{equation}
\left[1,1,N-1\right].
\end{equation}
The corresponding left eigenvectors are
\begin{equation}
\left[\begin{array}{c}
-i\omega_{p}\\
\omega_{1}\\
\omega_{2}\\
\vdots\\
\omega_{N-1}\\
\omega_{N}
\end{array}\right];\left[\begin{array}{c}
i\omega_{p}\\
\omega_{1}\\
\omega_{2}\\
\vdots\\
\omega_{N-1}\\
\omega_{N}
\end{array}\right];\left[\begin{array}{c}
0\\
1/\omega_{1}\\
0\\
\vdots\\
0\\
-1/\omega_{N}
\end{array}\right],\left[\begin{array}{c}
0\\
0\\
1/\omega_{2}\\
\vdots\\
0\\
-1/\omega_{N}
\end{array}\right],\dots,\left[\begin{array}{c}
0\\
0\\
0\\
\vdots\\
1/\omega_{N-1}\\
-1/\omega_{N}
\end{array}\right].
\end{equation}

The last $N-1$ eigenvectors for eigenvalue $i\omega=0$ are real
and do not depend on time. The first two eigenvectors due to eigenvalues
$i\omega=\pm i\omega_{p}$ are conjugate to each other. They can be
used to construct two real, orthogonal real solutions following Sec.~(\ref{subsec:appendix-skew-hermitian-real-solutions}).
The results are
\begin{equation}
\mathbf{a}\cos\omega_{p}t-\mathbf{b}\sin\omega_{p}t\quad\text{and}\quad\mathbf{a}\sin\omega_{p}t+\mathbf{b}\cos\omega_{p}t
\end{equation}
where 
\begin{equation}
\mathbf{a}=\left[\begin{array}{c}
0\\
\omega_{1}\\
\omega_{2}\\
\vdots\\
\omega_{N-1}\\
\omega_{N}
\end{array}\right]\text{ and }\mathbf{b}=\left[\begin{array}{c}
\omega_{p}\\
0\\
0\\
\vdots\\
0\\
0
\end{array}\right].
\end{equation}

\subsection{\label{subsec:appendix-exact-perp}Solving the Perpendicular Sub-System}

The perpendicular system evolves $\left(\mathbf{E}_{\perp};\mathbf{u}_{s\perp}\right)$
and writes

\begin{equation}
\frac{\partial}{\partial t}\left[\begin{array}{c}
\tilde{\mathbf{E}}_{\perp1}\\
\tilde{\mathbf{E}}_{\perp2}\\
\tilde{\mathbf{J}}_{1\perp1}\\
\tilde{\mathbf{J}}_{1\perp2}\\
\tilde{\mathbf{J}}_{2\perp1}\\
\tilde{\mathbf{J}}_{2\perp2}\\
\vdots\\
\tilde{\mathbf{J}}_{N\perp1}\\
\tilde{\mathbf{J}}_{N\perp2}
\end{array}\right]=\left[\begin{array}{ccccccccc}
0 & 0 & -\omega_{1} & 0 & -\omega_{1} & 0 & \cdots & -\omega_{N} & 0\\
0 & 0 & 0 & -\omega_{1} & 0 & -\omega_{1} & \cdots & 0 & -\omega_{N}\\
\omega_{1} & 0 & 0 & \Omega_{1} & 0 & 0 & \cdots & 0 & 0\\
0 & \omega_{1} & -\Omega_{1} & 0 & 0 & 0 & \cdots & 0 & 0\\
\omega_{2} & 0 & 0 & 0 & 0 & \Omega_{2} & \cdots & 0 & 0\\
0 & \omega_{2} & 0 & 0 & -\Omega_{2} & 0 & \cdots & 0 & 0\\
\vdots & \vdots & \vdots & \vdots & \vdots & \vdots & \ddots & \vdots & \vdots\\
\omega_{N} & 0 & 0 & 0 & 0 & 0 & \cdots & 0 & \Omega_{N}\\
0 & \omega_{N} & 0 & 0 & 0 & 0 & \cdots & -\Omega_{N} & 0
\end{array}\right]\cdot\left[\begin{array}{c}
\tilde{\mathbf{E}}_{\perp1}\\
\tilde{\mathbf{E}}_{\perp2}\\
\tilde{\mathbf{J}}_{1\perp1}\\
\tilde{\mathbf{J}}_{1\perp2}\\
\tilde{\mathbf{J}}_{2\perp1}\\
\tilde{\mathbf{J}}_{2\perp2}\\
\vdots\\
\tilde{\mathbf{J}}_{N\perp1}\\
\tilde{\mathbf{J}}_{N\perp2}
\end{array}\right].\label{eq:dE-dJ-perp-full}
\end{equation}
Here the subscripts $_{1}$ and $_{2}$ represent two orthogonal directions
that form a right-handed coordinate when combined with the background
magnetic field direction (along direction ``3'').

Note that the coefficient matrix of Eq.~(\ref{eq:dE-dJ-perp-full})
is a $2\left(N+1\right)\times2\left(N+1\right)$ real, skew-symmetric
matrix. Its nonzero eigenvalues are purely imaginary and come in pairs
$\pm i\omega$ with conjugate eigenvectors. In total, the system has
$2\left(N+1\right)$ eigenvalues and eigenvectors. However, to simplify
the problem, we may consider the two components of a perpendicular
vector as the real and imaginary parts of a complex vector, the equation
above can be rearranged into a more compact form Eq.~(\ref{eq:dE-dJ-perp-matrix-form}).
The coefficient matrix $\mathbf{M}_{\perp}$ now is a $\left(N+1\right)\times\left(N+1\right)$
skew-Hermitian matrix and has $N+1$ complex eigenvectors. The real
and imaginary parts of the eigenvectors serve as the 1st and 2nd components
of the perpendicular vectors, as we will see below.

\subsubsection{Eigenvalues}

Consider the eigenstructure of the perpendicular problem Eq.~(\ref{eq:dE-dJ-perp-matrix-form}),
\begin{equation}
0=\left(\mathbf{M}_{\perp}-i\omega\hat{\mathbf{I}}\right)\cdot\mathbf{q}_{\perp}=\left[\begin{array}{ccccc}
-i\omega & -\omega_{1} & -\omega_{2} & \cdots & -\omega_{N}\\
\omega_{1} & -i\,\left(\omega+\Omega_{1}\right) & 0 & 0 & 0\\
\omega_{2} & 0 & -i\,\left(\omega+\Omega_{2}\right) & 0 & 0\\
\vdots & 0 & 0 & \ddots & 0\\
\omega_{N} & 0 & 0 & 0 & -i\,\left(\omega+\Omega_{N}\right)
\end{array}\right]\cdot\mathbf{q}_{\perp}.\label{eq:exact-perp-eigenstructure}
\end{equation}
The first row gives
\begin{equation}
-i\omega\tilde{\mathbf{E}}_{\perp}=\sum_{s}\omega_{s}\tilde{\mathbf{J}}_{s\perp}\label{eq:exact-perp-relation-1}
\end{equation}
while the remaining rows simultaneously give 
\begin{equation}
\tilde{\mathbf{E}}_{\perp}\omega_{s}-i\,\left(\omega+\Omega_{s}\right)\tilde{\mathbf{J}}_{\perp s}=0\text{, where \ensuremath{s=1,\dots,N}}
\end{equation}
\begin{equation}
\Rightarrow\tilde{\mathbf{J}}_{\perp s}=\tilde{\mathbf{E}}_{\perp}\frac{\omega_{s}}{i\left(\omega+\Omega_{s}\right)}.\label{eq:exact-perp-relation-2}
\end{equation}
Substitute Eq.~(\ref{eq:exact-perp-relation-2}) into Eq.~(\ref{eq:exact-perp-relation-1})
leads to
\begin{equation}
-i\omega\tilde{\mathbf{E}}_{\perp}=\sum_{s}\omega_{s}\frac{\omega_{s}}{i\left(\omega+\Omega_{s}\right)}\tilde{\mathbf{E}}_{\perp}=-i\sum_{s}\frac{\omega_{s}^{2}}{\omega+\Omega_{s}}\tilde{\mathbf{E}}_{\perp}
\end{equation}
or 
\begin{equation}
\tilde{\mathbf{E}}_{\perp}\left(\omega-\sum_{s}\frac{\omega_{s}^{2}}{\omega+\Omega_{s}}\right)=0\label{eq:appendix-exact-perp-eigenvalues}
\end{equation}
which has nontrivial solution of $\tilde{\mathbf{E}}_{\perp}$ when
and only when Eq.~(\ref{eq:exact-perp-eigenvalue}) is satisfied.
Since $\mathbf{M}_{\perp}$ is skew-Hermitian and $i\mathbf{M}_{\perp}$
is Hermitian, solving the equation above gives $N+1$ real eigenvalues
$\omega$ for $i\mathbf{M}_{\perp}$ and correspondingly $N+1$ imaginary
eigenvalues $\lambda=i\omega$ for $\mathbf{M}_{\perp}$.

Eq.~(\ref{eq:exact-perp-eigenvalue}) can be expanded as a $N+1$
order polynomial and solved with any root finder. For three or fewer
species, analytic formulae exists for the roots. For more species,
we may find the roots as eigenvalues of a companion matrix, or using
an iterative root finder.

\subsubsection{Eigenvectors}

As indicated by Eq.~(\ref{eq:exact-perp-relation-2}), for each eigenvalue
$i\omega$, there are two complex eigenvectors
\begin{equation}
\left[\begin{array}{c}
i\\
\frac{\omega_{1}}{\omega+\Omega_{1}}\\
\frac{\omega_{2}}{\omega+\Omega_{2}}\\
\frac{\omega_{3}}{\omega+\Omega_{3}}\\
\frac{\omega_{4}}{\omega+\Omega_{4}}
\end{array}\right]\text{ and }\left[\begin{array}{c}
1\\
-i\frac{\omega_{1}}{\omega+\Omega_{1}}\\
-i\frac{\omega_{2}}{\omega+\Omega_{2}}\\
-i\frac{\omega_{3}}{\omega+\Omega_{3}}\\
-i\frac{\omega_{4}}{\omega+\Omega_{4}}
\end{array}\right].
\end{equation}

\subsubsection{Real solution bases}

From these two complex eigenvectors, we can determine two time-dependent
real, orthogonal solution bases following the recipe in Sec.~\ref{subsec:appendix-skew-hermitian-real-solutions}.
To determine the first real solution, we first write down the full,
complex eigenvector of $\lambda=i\omega$ as
\begin{equation}
\mathbf{v}=\mathbf{a}+i\mathbf{b}\text{, where }\mathbf{a}=\left[\begin{array}{c}
0\\
\frac{\omega_{1}}{\omega+\Omega_{1}}\\
\frac{\omega_{2}}{\omega+\Omega_{2}}\\
\frac{\omega_{3}}{\omega+\Omega_{3}}\\
\frac{\omega_{4}}{\omega+\Omega_{4}}
\end{array}\right],\mathbf{b}=\left[\begin{array}{c}
1\\
0\\
0\\
0\\
0
\end{array}\right],
\end{equation}
then the time-dependent solution for the perpendicular problem is 

\begin{align}
\mathbf{v}e^{i\omega t} & =\left(\mathbf{a}+i\mathbf{b}\right)\left(\cos\omega t+i\sin\omega t\right)\nonumber \\
 & =\left(\mathbf{a}\cos\omega t-\mathbf{b}\sin\omega t\right)+i\left(\mathbf{a}\sin\omega t+\mathbf{b}\cos\omega t\right)\nonumber \\
 & =\left[\begin{array}{c}
-\sin\omega t\\
\frac{\omega_{1}}{\omega+\Omega_{1}}\cos\omega t\\
\frac{\omega_{2}}{\omega+\Omega_{2}}\cos\omega t\\
\frac{\omega_{3}}{\omega+\Omega_{3}}\cos\omega t\\
\frac{\omega_{4}}{\omega+\Omega_{4}}\cos\omega t
\end{array}\right]+i\left[\begin{array}{c}
\cos\omega t\\
\frac{\omega_{1}}{\omega+\Omega_{1}}\sin\omega t\\
\frac{\omega_{2}}{\omega+\Omega_{2}}\sin\omega t\\
\frac{\omega_{3}}{\omega+\Omega_{3}}\sin\omega t\\
\frac{\omega_{4}}{\omega+\Omega_{4}}\sin\omega t
\end{array}\right].
\end{align}
The real and imaginary parts in each row are the first and second
components of a same perpendicular vector $\tilde{\mathbf{E}}_{\perp}$
or $\tilde{\mathbf{J}}_{s\perp}$. Thus the time-dependent real solution
is
\begin{equation}
\left[\begin{array}{c}
\tilde{E}_{1}\\
\tilde{E}_{2}\\
\vdots\\
J_{s1}\\
\tilde{J}_{s2}\\
\vdots
\end{array}\right]=\left[\begin{array}{c}
-\sin\omega t\\
\cos\omega t\\
\vdots\\
\frac{\omega_{s}}{\omega+\Omega_{s}}\cos\omega t\\
\frac{\omega_{s}}{\omega+\Omega_{s}}\sin\omega t\\
\vdots
\end{array}\right].
\end{equation}

Similarly, the second time-dependent real solution can be determined
by 
\begin{equation}
\mathbf{v}=\mathbf{a}+i\mathbf{b}\text{, where }\mathbf{a}=\left[\begin{array}{c}
1\\
0\\
0\\
0\\
0
\end{array}\right],\mathbf{b}=-\left[\begin{array}{c}
0\\
\frac{\omega_{1}}{\omega+\Omega_{1}}\\
\frac{\omega_{2}}{\omega+\Omega_{2}}\\
\frac{\omega_{3}}{\omega+\Omega_{3}}\\
\frac{\omega_{4}}{\omega+\Omega_{4}}
\end{array}\right],
\end{equation}

\begin{align}
\Rightarrow\mathbf{v}e^{i\omega t} & =\left(\mathbf{a}+i\mathbf{b}\right)\left(\cos\omega t+i\sin\omega t\right)\nonumber \\
 & =\left(\mathbf{a}\cos\omega t-\mathbf{b}\sin\omega t\right)+i\left(\mathbf{a}\sin\omega t+\mathbf{b}\cos\omega t\right)\nonumber \\
 & =\left[\begin{array}{c}
\cos\omega t\\
\frac{\omega_{1}}{\omega+\Omega_{1}}\sin\omega t\\
\frac{\omega_{2}}{\omega+\Omega_{2}}\sin\omega t\\
\frac{\omega_{3}}{\omega+\Omega_{3}}\sin\omega t\\
\frac{\omega_{4}}{\omega+\Omega_{4}}\sin\omega t
\end{array}\right]+i\left[\begin{array}{c}
\sin\omega t\\
-\frac{\omega_{1}}{\omega+\Omega_{1}}\cos\omega t\\
-\frac{\omega_{2}}{\omega+\Omega_{2}}\cos\omega t\\
-\frac{\omega_{3}}{\omega+\Omega_{3}}\cos\omega t\\
-\frac{\omega_{4}}{\omega+\Omega_{4}}\cos\omega t
\end{array}\right].
\end{align}
The second real solution is then
\begin{equation}
\left[\begin{array}{c}
\tilde{E}_{1}\\
\tilde{E}_{2}\\
\vdots\\
J_{s1}\\
\tilde{J}_{s2}\\
\vdots
\end{array}\right]=\left[\begin{array}{c}
\cos\omega t\\
\sin\omega t\\
\vdots\\
\frac{\omega_{s}}{\omega+\Omega_{s}}\sin\omega t\\
-\frac{\omega_{s}}{\omega+\Omega_{s}}\cos\omega t\\
\vdots
\end{array}\right].
\end{equation}

Now we find two real, orthogonal solutions for $\lambda=i\omega$
of different polarizations as in Eq.~(\ref{eq:exact-perp-real-solutions}).
In total, we recover all $2\times\left(N+1\right)$ real, orthogonal
solutions to the $2\left(N+1\right)$ order Eq.~(\ref{eq:dE-dJ-perp-full}).

\section{\label{subsec:appendix-implicit-direct}Direct Calculation of the
Locally Implicit Scheme}

The linear, constant-coefficient ODEs Eqns.\,(\ref{eq:dJdt_dEdt_implicit-with-collision})
can be solved directly instead of through matrix inversion, as described
in Ref.\,\citep{Smithe2007a}. This usually leads to faster computation.
Here, we give a straightforward derivation, and fix a few minor mistakes
in Ref.\,\citep{Smithe2007a}. Note that we do not consider conllisions,
while Ref.\,\citep{Smithe2007a} did.

We start by noting that the general problem 
\begin{equation}
\mathbf{A}=\mathbf{R}+\mathbf{A}\times\mathbf{B},\label{eq:appendix-implicit-direct-useful-A_R+AxB}
\end{equation}
where $\mathbf{R}$ and $\mathbf{B}$ are knowns, has the solution
\begin{equation}
\mathbf{A}=\frac{\mathbf{R}+\mathbf{B}\mathbf{B}\cdot\mathbf{R}-\mathbf{B}\times\mathbf{R}}{1+B^{2}}=\frac{\mathbf{R}+B^{2}\mathbf{b}\mathbf{b}\cdot\mathbf{R}-B\mathbf{b}\times\mathbf{R}}{1+B^{2}}.\label{eq:appendix-implicit-direct-useful-A_sol}
\end{equation}
A relevant problem 
\begin{equation}
\mathbf{C}=\mathbf{R}+\mathbf{C}\times\mathbf{B}+\xi\mathbf{b}\mathbf{b}\cdot\mathbf{C},\label{eq:appendix-implicit-direct-useful-C_R+CxB+xiC_par}
\end{equation}
where $\xi\neq1$ and $\mathbf{b}$ is the unit vector along $\mathbf{B}$,
has the solution 
\begin{equation}
\mathbf{C}=\frac{\mathbf{R}+\mathbf{B}\mathbf{B}\cdot\mathbf{R}-\mathbf{B}\times\mathbf{R}}{1+B^{2}}+\frac{\xi}{1-\xi}\mathbf{b}\mathbf{b}\cdot\mathbf{R}.\label{eq:appendix-implicit-direct-useful-C_sol}
\end{equation}

The first line of Eqns.\,(\ref{eq:dJdt_dEdt_implicit-with-collision})
can be re-arranged into the form of Eqn.\,(\ref{eq:appendix-implicit-direct-useful-A_R+AxB}),
\begin{equation}
\bar{\mathbf{J}}_{s}=\left(\mathbf{J}_{s}^{n}+\frac{\varepsilon_{0}\omega_{s}^{2}\Delta t}{2}\bar{\mathbf{E}}\right)+\bar{\mathbf{J}}_{s}\times\left(\frac{\Omega_{s}\Delta t}{2}\mathbf{b}\right),
\end{equation}
thus has the solution 
\begin{equation}
\bar{\mathbf{J}}_{s}=\left(1+\frac{\Omega_{s}^{2}\Delta t^{2}}{4}\right)^{-1}\left(\mathbf{J}_{s}^{*}+\frac{\Omega_{s}^{2}\Delta t^{2}}{4}\mathbf{b}\mathbf{b}\cdot\mathbf{J}_{s}^{*}-\mathbf{b}\times\frac{\Omega_{s}\Delta t}{2}\mathbf{J}_{s}^{*}\right),\label{eq:appendix-implicit-direct-useful-J_sol}
\end{equation}
where 
\begin{equation}
\mathbf{J}_{s}^{*}\equiv\mathbf{J}_{s}^{n}+\frac{\varepsilon_{0}\omega_{s}^{2}\Delta t}{2}\bar{\mathbf{E}}.\label{eq:appendix-implicit-direct-useful-J_tilde}
\end{equation}
Note that Eqn.\,(15) of Ref.\,\citep{Smithe2007a} corresponds to
Eqn.\,((\ref{eq:appendix-implicit-direct-useful-J_sol})) above,
but it misses a leading coefficient $\left(1+\Omega_{s}^{2}\Delta t^{2}/4\right)^{-1}$
and the dimensionality is not correct.

Substituting Eqn.\,(\ref{eq:appendix-implicit-direct-useful-J_sol})
back into the second line of Eqns.\,(\ref{eq:dJdt_dEdt_implicit-with-collision})
yields an equation of $\bar{\mathbf{E}}$ in the form of Eqn.\,(\ref{eq:appendix-implicit-direct-useful-C_R+CxB+xiC_par}):
\begin{eqnarray}
\mathbf{F}^{n+1/2} & = & \frac{\mathbf{F}^{n}+\frac{1}{2}\mathbf{K}}{1+\frac{1}{4}\omega_{0}^{2}}+\frac{1}{1+\frac{1}{4}\omega_{0}^{2}}\frac{\delta}{8}\mathbf{b}\times\mathbf{F}^{n+1/2}\nonumber \\
 &  & -\frac{1}{1+\frac{1}{4}\omega_{0}^{2}}\frac{\gamma^{2}}{16}\mathbf{b}\mathbf{b}\cdot\mathbf{F}^{n+1/2},\label{eq:appendix-implicit-direct-useful-F_n+1/2}
\end{eqnarray}
where $\mathbf{F}\equiv\varepsilon_{0}\mathbf{E}$, and 
\begin{eqnarray}
\mathbf{K} & \equiv & -\Delta t\sum_{s}\left[\left(1+\frac{\Omega_{s}^{2}\Delta t^{2}}{4}\right)^{-1}\right.\nonumber \\
 &  & \left.\left(\mathbf{J}_{s}^{n}+\frac{\Omega_{s}^{2}\Delta t^{2}}{4}\mathbf{b}\mathbf{b}\cdot\mathbf{J}_{s}^{n}-\mathbf{b}\times\frac{\Omega_{s}\Delta t}{2}\mathbf{J}_{s}^{n}\right)\right].
\end{eqnarray}
Following Eqn.\,(\ref{eq:appendix-implicit-direct-useful-C_sol})
and after some rather tedious algebraic re-arrangement, we obtain
\begin{eqnarray}
\bar{\mathbf{F}} & = & +\frac{1}{1+\frac{1}{4}\omega_{0}^{2}+\frac{1}{64}\Delta^{2}}\left(\mathbf{F}^{n}+\frac{1}{2}\mathbf{K}\right)\nonumber \\
 &  & \frac{\frac{1}{64}\Delta^{2}-\frac{1}{16}\gamma^{2}}{\left(1+\frac{1}{4}\omega_{0}^{2}+\frac{1}{64}\Delta^{2}\right)\left(1+\frac{1}{4}\omega_{0}^{2}+\frac{1}{16}\gamma^{2}\right)}\mathbf{b}\mathbf{b}\cdot\left(\mathbf{F}^{n}+\frac{1}{2}\mathbf{K}\right)\nonumber \\
 &  & +\frac{\frac{1}{8}\delta}{\left(1+\frac{1}{4}\omega_{0}^{2}+\frac{1}{64}\Delta^{2}\right)\left(1+\frac{1}{4}\omega_{0}^{2}\right)}\mathbf{b}\times\left(\mathbf{F}^{n}+\frac{1}{2}\mathbf{K}\right).\label{eq:appendix-implicit-direct-useful-F_sol}
\end{eqnarray}
Here, we use the following notations modified from Eqn.\,(11) of
Ref.\,\citep{Smithe2007a}: 
\begin{eqnarray}
\omega_{0}^{2} & \equiv & \sum_{s}\frac{\omega_{s}^{2}\Delta t^{2}}{1+\Omega_{s}^{2}\Delta t^{2}/4},\nonumber \\
\gamma^{2} & = & \sum_{s}\frac{\omega_{s}^{2}\Omega_{s}^{2}\Delta t^{4}}{1+\Omega_{s}^{2}\Delta t^{2}/4},\delta=\sum_{s}\frac{\omega_{s}^{2}\Omega_{s}\Delta t^{3}}{1+\Omega_{s}^{2}\Delta t^{2}/4},\label{eq:appendix-implicit-direct-useful-useful-identities}\\
\Delta & \equiv & \frac{\delta^{2}}{1+\omega_{0}^{2}}.\nonumber 
\end{eqnarray}
This result is consistent with Eqn.\,(13) of Ref.\,\citep{Smithe2007a}
though the latter missed a few terms.

In the actual implementation, Eqn.\,(\ref{eq:appendix-implicit-direct-useful-F_sol})
is computed first to get $\bar{\mathbf{E}}$, which is substituted
in Eqn.\,(\ref{eq:appendix-implicit-direct-useful-J_sol}) to compute
$\bar{\mathbf{J}}_{s}$ for each species. The final updated currents
and electric fields are then determined by $\mathbf{J}_{s}^{n+1}=2\bar{\mathbf{J}}_{s}-\mathbf{J}_{s}^{n}$
and $\mathbf{E}^{n+1}=2\bar{\mathbf{E}}-\mathbf{E}^{n}$. In an informal
two-fluid five-moment test, the exact source solution described here
is approximately $5\times$ times faster than the solution through
matrix inversion using external numerical package.

\section{\label{subsec:stability}Stability of the Locally Implict Scheme}

\subsection{Von Neumann Stability Analysis}

Following the Von Neumann analysis, we assume all quantities depend
on time as $e^{i\omega t}$:
\begin{align}
\mathbf{q}_{s}^{n+1} & =\mathbf{q}_{s}^{n}e^{i\omega\Delta t},\quad\mathbf{q}=\left[\mathbf{E};\mathbf{J}_{s}\right]^{T}.
\end{align}
The time-centered quantities are
\begin{align}
\bar{\mathbf{q}}_{s} & =\frac{\mathbf{q}_{s}^{n+1}+\mathbf{q}_{s}^{n}}{2}=\mathbf{q}_{s}^{n}\frac{1+e^{i\omega\Delta t}}{2}.
\end{align}
For simplicity, we consider the normalized quantities $\tilde{\mathbf{E}}$
and $\tilde{\mathbf{J}}_{s}$ as defined in Section.~\ref{sec:exact}
in this derivation.

First, considering only the plasma oscillation, i.e., neglecting the
cyclotron term $\mathbf{J}_{s}\times\boldsymbol{\Omega}_{s}$, the
locally implicit scheme gives
\begin{equation}
\begin{cases}
\bar{\tilde{\mathbf{E}}} & =\mathbf{\tilde{E}}^{n}-\frac{\Delta t}{2}\omega_{ps}\sum_{s}\bar{\mathbf{\tilde{\mathbf{J}}}}_{s}.\\
\bar{\mathbf{\tilde{\mathbf{J}}}}_{s} & =\mathbf{\tilde{\mathbf{J}}}_{s}^{n}+\frac{\Delta t\omega_{s}}{2}\bar{\tilde{\mathbf{E}}}
\end{cases}
\end{equation}
or
\begin{equation}
\mathbf{q}^{n}\frac{1+e^{i\omega\Delta t}}{2}=\mathbf{q}^{n}+\frac{\Delta t}{2}\mathbf{M}_{\parallel}\mathbf{q}^{n}\frac{1+e^{i\omega\Delta t}}{2},\label{eq:stability-para-0}
\end{equation}
where $\mathbf{q}=\left[\tilde{\mathbf{E}};\tilde{\mathbf{J}}_{s}\right]^{T}$
and the coefficient matrix is defined in Eqn.~\ref{eq:para-mat}.
Eqn.~(\ref{eq:stability-para-0}) can be rearranged to get
\begin{equation}
\mathbf{q}^{n}\frac{e^{i\omega\Delta t}-1}{2}=\frac{\Delta t}{2}\mathbf{M}_{\parallel}\mathbf{q}^{n}\frac{e^{i\omega\Delta t}+1}{2}\label{eq:stability-para-1}
\end{equation}
or
\begin{equation}
\mathbf{q}^{n}i\tan\left(\frac{\omega\Delta t}{2}\right)=\frac{\Delta t}{2}\mathbf{M}_{\parallel}\mathbf{q}^{n}.\label{eq:stability-para-2}
\end{equation}
Here, we used the relation
\begin{equation}
1-e^{-i\theta}=2i\sin\frac{\theta}{2}e^{-i\theta/2}\quad{\rm and}\quad1+e^{-i\theta}=2\cos\frac{\theta}{2}e^{-i\theta/2}.
\end{equation}
Recall that $\mathbf{M}_{\parallel}$ has nonzero eigenvalues $\pm i\omega_{p}$,
therefore Eqn.~(\ref{eq:stability-para-2}) gives the stability criterion
for plasma oscillation
\begin{equation}
\left(\frac{2}{\Delta t}\right)^{2}\tan^{2}\left(\frac{\omega\Delta t}{2}\right)=\omega_{p}^{2}.
\end{equation}
This equation has only real solutions for $\omega$ thus eliminates
the possibility of numerical instability.

Similarly, we may compute the stability criteria for the perpendicular
problem. The results will be of the form
\begin{equation}
\left(\frac{2}{\Delta t}\right)^{2}\tan^{2}\left(\frac{\omega\Delta t}{2}\right)=\omega_{\perp}^{2}
\end{equation}
where $i\omega_{\perp}$ is an eigenvalue of the matrix $\mathbf{M}_{\perp}$
defined in Eqn.~(\ref{eq:perp-mat}). Again, the time-step is not
restricted by $\omega_{\perp}$, which contains both plasma and cyclotron
frequencies.

\subsection{Properties of the Implicit Midpoint Method}

The locally implicit scheme is essentially an implicit midpoint method.
Thus it is useful to understand the general properties of the method.
Consider the initial value problem of an ODE,
\[
y^{\prime}(t)=f(t,y(t)),\quad y(t_{0})=y_{0}.
\]
The implicit midpoint method is given by

\begin{equation}
y_{n+1}=y_{n}+hf\left(t_{n}+\frac{h}{2},\,\frac{y_{n}+y_{n+1}}{2}\right).\label{eq:method-implicit-midpoint-a}
\end{equation}
It can be written as an implicit Runge-Kutta method
\begin{align}
k & =f\left(t_{n}+\frac{h}{2},\,y_{n}+\frac{h}{2}k\right)\nonumber \\
y_{n+1} & =y_{n}+hk\label{eq:method-implicit-midpoint-b}
\end{align}
which contains the implicit Euler method with step size $h/2$ as
its first part. We may also write the method as
\begin{align}
\bar{y} & =y_{n}+\frac{h}{2}f\left(t_{n}+\frac{h}{2},\,\bar{y}\right)\nonumber \\
y_{n+1} & =y_{n}+hf\left(t_{n}+\frac{h}{2},\,\bar{y}\right)\label{eq:method-implicit-midpoint-c}
\end{align}
since
\begin{equation}
y_{n+1}-2\bar{y}=y_{n}-2y_{n}=-y_{n}\Leftrightarrow\bar{y}=\frac{y_{n}+y_{n+1}}{2}.
\end{equation}
An additional observation is that the second step in Eqn.~(\ref{eq:method-implicit-midpoint-c})
can then be replaced by $y_{n+1}=2\bar{y}-y_{n}$. This simplifies
the implementaiton and is used in our code.

The implicit midpoint method has local truncation error of order $O\left(h^{3}\right)$
hence global error of order $O\left(h^{2}\right)$. For a problem
$y=e^{\lambda t}$, the stability region of the method is the entire
half plane with $Im\left(\lambda\right)h\leq0$, thus the method is
unconditionally stable for nongrowing problems. For an purely oscillatory
problem, like our source update equations, $\lambda$ lies right on
the border of the stability region (the imaginary axis), indicating
exact preservation of oscillation magnitude.
\begin{figure}
\begin{centering}
\includegraphics[width=0.4\textwidth]{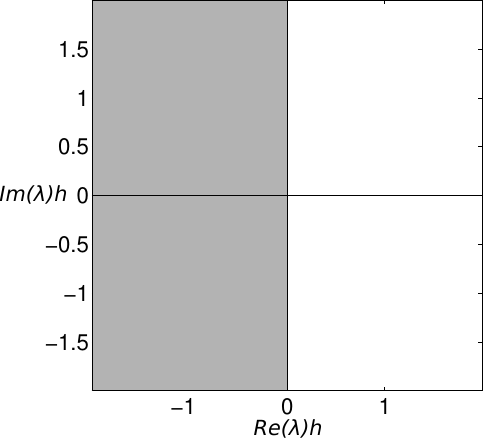}
\par\end{centering}
\caption{The gray shadows represent the stability region of the implicit midpoint
method on the complex plane. A purely oscillatory problem lies on the imaginary
axis, which is exactly the boundary of the stability region. Therefore, when applied
to a purely oscillatory problem, the locally midpoint method conserves the amplitude,
i.e., no damping or amplifying.}

\end{figure}

\section{The Eigensystem of the Ten-Moment Model}

To determine the eigensystem of the homogeneous part of the ten-moment
equations we first write, in one-dimension, the left-hand side of
Eqns.\,(\ref{eq:drhodt_full}), (\ref{eq:drhovdt_full}) and (\ref{eq:dPdt_full}),
in the quasilinear form 
\begin{equation}
\partial_{t}\mathbf{v}+\mathbf{A}\partial_{1}\mathbf{v}=0\label{eq:qlForm}
\end{equation}
where $\mathbf{v}$ is the vector of primitive variables and $\mathbf{A}$
is a matrix of coefficients. For the ten-moment system we have 
\begin{align}
\mathbf{v}=\left[\begin{matrix}\rho,u_{1},u_{2},u_{3},P_{11},P_{12},P_{13},P_{22},P_{23},P_{33}\end{matrix}\right]^{T}
\end{align}
where $\rho\equiv mn$ and 
\begin{align}
\mathbf{A}=\left[\begin{matrix}u_{1} & \rho & 0 & 0 & 0 & 0 & 0 & 0 & 0 & 0\\
0 & u_{1} & 0 & 0 & 1/\rho & 0 & 0 & 0 & 0 & 0\\
0 & 0 & u_{1} & 0 & 0 & 1/\rho & 0 & 0 & 0 & 0\\
0 & 0 & 0 & u_{1} & 0 & 0 & 1/\rho & 0 & 0 & 0\\
0 & 3P_{11} & 0 & 0 & u_{1} & 0 & 0 & 0 & 0 & 0\\
0 & 2P_{12} & P_{11} & 0 & 0 & u_{1} & 0 & 0 & 0 & 0\\
0 & 2P_{13} & 0 & P_{11} & 0 & 0 & u_{1} & 0 & 0 & 0\\
0 & P_{22} & 2P_{12} & 0 & 0 & 0 & 0 & u_{1} & 0 & 0\\
0 & P_{23} & P_{13} & P_{12} & 0 & 0 & 0 & 0 & u_{1} & 0\\
0 & P_{33} & 0 & 2P_{13} & 0 & 0 & 0 & 0 & 0 & u_{1}
\end{matrix}\right]
\end{align}
The eigensystem of this matrix can be easily obtained either by hand
or a computer algebra package. The results are described below.

The eigenvalues of the system are given by 
\begin{align}
\lambda^{1,2} & =u_{1}-\sqrt{P_{11}/\rho}\\
\lambda^{3,4} & =u_{1}+\sqrt{P_{11}/\rho}\\
\lambda^{5} & =u_{1}-\sqrt{3P_{11}/\rho}\\
\lambda^{6} & =u_{1}+\sqrt{3P_{11}/\rho}\\
\lambda^{7,8,9,10} & =u_{1}
\end{align}
To maintain hyperbolicity we must hence have $\rho>0$ and $P_{11}>0$.
In multiple dimensions, in general, the diagonal elements of the pressure
tensor must be positive. When $P_{11}=0$ the system reduces to the
cold fluid equations which is known to be rank deficient and hence
not hyperbolic as usually understood\footnote{For hyperbolicity the matrix $A$ must posses real eigenvalues and
a complete set of linearly independent right eigenvectors. For the
cold fluid system we only have a single eigenvalue (the fluid velocity)
and a single eigenvector. This can lead to generalized solutions like
delta shocks.}. Also notice that the eigenvalues do not include the usual fluid
sound-speed $c_{s}=\sqrt{5p/3\rho}$ but instead have two different
propagation speeds $c_{1}=\sqrt{P_{11}/\rho}$ and $c_{2}=\sqrt{3P_{11}/\rho}$.
This is because the (neutral) ten-moment system does not go to the
correct limit of Euler equations in the absence of collisions. In
fact, it is collisions that drive the pressure tensor to isotropy,
but do not appear in the homogenous ten-moment system.

The right eigenvectors (column vectors) are given below. 
\begin{align}
\mathbf{r}^{1,3}=\left[\begin{matrix}0\\
0\\
\mp c_{1}\\
0\\
0\\
P_{11}\\
0\\
2P_{12}\\
P_{13}\\
0
\end{matrix}\right]\quad\mathbf{r}^{2,4}=\left[\begin{matrix}0\\
0\\
0\\
\mp c_{1}\\
0\\
0\\
P_{11}\\
0\\
P_{12}\\
2P_{13}
\end{matrix}\right]
\end{align}
and 
\begin{align}
\mathbf{r}^{5,6}=\left[\begin{matrix}\rho P_{11}\\
\mp c_{2}P_{11}\\
\mp c_{2}P_{12}\\
\mp c_{2}P_{13}\\
3P_{11}^{2}\\
3P_{11}P_{12}\\
3P_{11}P_{13}\\
P_{11}P_{22}+2P_{12}^{2}\\
P_{11}P_{23}+2P_{12}P_{13}\\
P_{11}P_{33}+2P_{13}^{2}
\end{matrix}\right]
\end{align}
and 
\begin{align}
\mathbf{r}^{7}=\left[\begin{matrix}1\\
0\\
0\\
0\\
0\\
0\\
0\\
0\\
0\\
0
\end{matrix}\right]\quad\mathbf{r}^{8}=\left[\begin{matrix}0\\
0\\
0\\
0\\
0\\
0\\
0\\
1\\
0\\
0
\end{matrix}\right]\quad\mathbf{r}^{9}=\left[\begin{matrix}0\\
0\\
0\\
0\\
0\\
0\\
0\\
0\\
1\\
0
\end{matrix}\right]\quad\mathbf{r}^{10}=\left[\begin{matrix}0\\
0\\
0\\
0\\
0\\
0\\
0\\
0\\
0\\
1
\end{matrix}\right]
\end{align}

We can now compute the left eigenvectors (row vectors) by inverting
the matrix with right eigenvectors stored as columns. This ensures
the normalization $\mathbf{l}^{p}\mathbf{r}^{k}=\delta^{pk}$, where
the $\mathbf{l}^{p}$ are the left eigenvectors. On performing the
inversion we have 
\begin{align}
\mathbf{l}^{1,3} & =\left[\begin{matrix}0 & \pm\dfrac{P_{12}}{2c_{1}P_{11}} & \mp\dfrac{1}{2c_{1}} & 0 & -\dfrac{P_{12}}{2P_{11}^{2}} & \dfrac{1}{2P_{11}} & 0 & 0 & 0 & 0\end{matrix}\right]\\
\mathbf{l}^{2,4} & =\left[\begin{matrix}0 & \pm\dfrac{P_{13}}{2c_{1}P_{11}} & 0 & \mp\dfrac{1}{2c_{1}} & -\dfrac{P_{13}}{2P_{11}^{2}} & 0 & \dfrac{1}{2P_{11}} & 0 & 0 & 0\end{matrix}\right]
\end{align}
and 
\begin{align}
\mathbf{l}^{5,6}=\left[\begin{matrix}0 & \mp\dfrac{1}{2c_{2}P_{11}} & 0 & 0 & \dfrac{1}{6P_{11}^{2}} & 0 & 0 & 0 & 0 & 0\end{matrix}\right]
\end{align}
and 
\begin{align}
\mathbf{l}^{7} & =\left[\begin{matrix}1 & 0 & 0 & 0 & -\dfrac{1}{3c_{1}^{2}} & 0 & 0 & 0 & 0 & 0\end{matrix}\right]\\
\mathbf{l}^{8} & =\left[\begin{matrix}0 & 0 & 0 & 0 & \dfrac{4P_{12}^{2}-P_{11}P_{22}}{3P_{11}^{2}} & -\dfrac{2P_{12}}{P_{11}} & 0 & 1 & 0 & 0\end{matrix}\right]\\
\mathbf{l}^{9} & =\left[\begin{matrix}0 & 0 & 0 & 0 & \dfrac{4P_{12}P_{13}-P_{11}P_{23}}{3P_{11}^{2}} & -\dfrac{P_{13}}{P_{11}} & -\dfrac{P_{12}}{P_{11}} & 0 & 1 & 0\end{matrix}\right]\\
\mathbf{l}^{10} & =\left[\begin{matrix}0 & 0 & 0 & 0 & \dfrac{4P_{13}^{2}-P_{11}P_{33}}{3P_{11}^{2}} & 0 & -\dfrac{2P_{13}}{P_{11}} & 0 & 0 & 1\end{matrix}\right]
\end{align}

Most often, for numerical simulations, the eigensystem of the conservation
form of the homogeneous system is needed. This eigensystem is related
to the eigensystem of the quasilinear form derived above. To see this
consider a conservation law 
\begin{align}
\partial_{t}\mathbf{q}+\partial_{1}\mathbf{f}=0
\end{align}
where $\mathbf{f}=\mathbf{f}(\mathbf{q})$ is a flux function. Now
consider an invertible transformation $\mathbf{q}=\varphi(\mathbf{v})$.
This transforms the conservation law to 
\begin{align}
\partial_{t}\mathbf{v}+(\varphi')^{-1}\ D\mathbf{f}\ \varphi'\partial_{1}\mathbf{v}=0
\end{align}
where $\varphi'$ is the Jacobian matrix of the transformation and
$D\mathbf{f}\equiv\partial\mathbf{f}/\partial\mathbf{q}$ is the flux
Jacobian. Comparing this to Eqn.~\ref{eq:qlForm} we see that the
quasilinear matrix is related to the flux Jacobian by 
\begin{align}
\mathbf{A}=(\varphi')^{-1}\ D\mathbf{f}\ \varphi'
\end{align}
This clearly shows that the eigenvalues of the flux Jacobian are the
same as those of the quasilinear matrix while the right and left eigenvectors
can be computed using $\varphi'\mathbf{r}^{p}$ and $\mathbf{l}^{p}(\varphi')^{-1}$
respectively.

For the ten-moment system the required transformation is 
\begin{align}
\mathbf{q}=\varphi(\mathbf{v})=\left[\begin{matrix}\rho\\
\rho u_{1}\\
\rho u_{2}\\
\rho u_{3}\\
\rho u_{1}u_{1}+P_{11}\\
\rho u_{1}u_{2}+P_{12}\\
\rho u_{1}u_{3}+P_{13}\\
\rho u_{2}u_{2}+P_{22}\\
\rho u_{2}u_{3}+P_{23}\\
\rho u_{3}u_{3}+P_{33}
\end{matrix}\right]
\end{align}
For this transformation we have 
\begin{align}
\varphi'(\mathbf{v})=\left[\begin{matrix}1 & 0 & 0 & 0 & 0 & 0 & 0 & 0 & 0 & 0\\
u_{1} & \rho & 0 & 0 & 0 & 0 & 0 & 0 & 0 & 0\\
u_{2} & 0 & \rho & 0 & 0 & 0 & 0 & 0 & 0 & 0\\
u_{3} & 0 & 0 & \rho & 0 & 0 & 0 & 0 & 0 & 0\\
u_{1}u_{1} & 2\rho u_{1} & 0 & 0 & 1 & 0 & 0 & 0 & 0 & 0\\
u_{1}u_{2} & \rho u_{2} & \rho u_{1} & 0 & 0 & 1 & 0 & 0 & 0 & 0\\
u_{1}u_{3} & \rho u_{3} & 0 & \rho u_{1} & 0 & 0 & 1 & 0 & 0 & 0\\
u_{2}u_{2} & 0 & 2\rho u_{2} & 0 & 0 & 0 & 0 & 1 & 0 & 0\\
u_{2}u_{3} & 0 & \rho u_{3} & \rho u_{2} & 0 & 0 & 0 & 0 & 1 & 0\\
u_{3}u_{3} & 0 & 0 & 2\rho u_{3} & 0 & 0 & 0 & 0 & 0 & 1
\end{matrix}\right]
\end{align}
The inverse of the transformation Jacobian is 
\begin{align}
(\varphi')^{-1}=\left[\begin{matrix}1 & 0 & 0 & 0 & 0 & 0 & 0 & 0 & 0 & 0\\
-u_{1}/\rho & 1/\rho & 0 & 0 & 0 & 0 & 0 & 0 & 0 & 0\\
-u_{2}/\rho & 0 & 1/\rho & 0 & 0 & 0 & 0 & 0 & 0 & 0\\
-u_{3}/\rho & 0 & 0 & 1/\rho & 0 & 0 & 0 & 0 & 0 & 0\\
u_{1}u_{1} & -2u_{1} & 0 & 0 & 1 & 0 & 0 & 0 & 0 & 0\\
u_{1}u_{2} & -u_{2} & -u_{1} & 0 & 0 & 1 & 0 & 0 & 0 & 0\\
u_{1}u_{3} & -u_{3} & 0 & -u_{1} & 0 & 0 & 1 & 0 & 0 & 0\\
u_{2}u_{2} & 0 & -2u_{2} & 0 & 0 & 0 & 0 & 1 & 0 & 0\\
u_{2}u_{3} & 0 & -u_{3} & -u_{2} & 0 & 0 & 0 & 0 & 1 & 0\\
u_{3}u_{3} & 0 & 0 & -2u_{3} & 0 & 0 & 0 & 0 & 0 & 1
\end{matrix}\right]
\end{align}

\bibliographystyle{elsarticle-num}
\bibliography{Multi-Fluid}

\end{document}